\documentclass[preprint,12pt]{aastex}
\usepackage{natbib}
\usepackage{amsmath}

\newcommand{\beq}{\begin{equation}}
\newcommand{\eeq}{\end{equation}}
\newcommand{\dhd}{\phi}
\newcommand{\eemtwo}{EM_{21}}
\newcommand{\eemone}{EM_{1}}
\newcommand{\ttwo}{t_{21}}
\newcommand{\CIT}{\psi}
\newcommand{\DEM}{\phi}

\newcommand{\lae}{\lesssim}

\newcommand{\temp}{{\hat T}}
\newcommand{\energy}{{\hat E}}
\newcommand{\ti}{{\hat T}_{i}}
\newcommand{\Ti}{T_{i}}
\newcommand{\To}{T_{o}}
\newcommand{\frad}{f_{rad}}
\newcommand{\Deltati}{\Delta {\hat T}_{i}}
\newcommand{\fline}{{\hat F}_{i}}
\newcommand{\kline}{{\hat K}_{i}}
\newcommand{\laplace}{L}
\newcommand{\tow}{{\hat T}}
\newcommand{\tinitial}{{\hat T}}
\newcommand{\tb}{{\hat T}_{b}}
\newcommand{\power}{p}
\newcommand{\unitstep}{H}

\begin{document}
\title{Parametrizing Impulsive X-ray Heating with a Cumulative Initial-Temperature Distribution}
%\title{An Alternative Approach to Modeling X-ray Spectra of Impulsively Heated Gas that is 
%More Physically Informative than the Differential Emission Measure}
\author{Kenneth G. Gayley}
\affil{Department of Physics and Astronomy, University of Iowa, Iowa
City, IA 5 2242}

%nodem outline:  Parametrizing the differential initial-temperature distribution instead of the 
%                differential emission measure

\begin{abstract}
In collisional ionization equilibrium (CIE), the X-ray spectrum from a plasma depends on the distribution of 
emission measure over temperature (DEM).
Due to the well-known ill conditioning problem, no precisely
resolved DEM can be inverted directly from the spectrum, so often only a gross parametrization of the DEM is used
to approximate the data, in hopes that the parametrization can 
provide useful model-independent constraints on the heating process.
However, ill conditioning also introduces ambiguity into the various different parametrizations
that could approximate the data, which may spoil the perceived
advantages of model independence.
Thus, this paper instead suggests a single parametrization for both the 
heating mechanism and the X-ray sources, based on a model of impulsive heating followed by complete cooling.
This approach is similar to a ``cooling flow'' approach, but allows injection at multiple initial temperatures,
and applies even when the steady state is a distribution
of different shock strengths, as for a standing shock with a range of obliquities, or for embedded stochastic shocks
that are only steady in a statistical sense.
This produces an alternative parametrization for X-ray spectra that is especially streamlined for 
higher density plasmas with efficient radiative
cooling, and provides internal consistency checks on the assumption of impulsive heating followed by complete cooling.
The result is no longer model independent, but the results are more directly interpretable in terms of useful physical
constraints on the impulsive heating distribution.

\end{abstract}

\section{Introduction}

Recent advances in X-ray spectroscopy have allowed high resolution, high signal-to-noise spectra to be used to
constrain the nature of hot astrophysical plasmas.
Yet even with modern data, the thermal bremsstrahlung intensity and line fluxes in each energy bin in a spectrum
receive contribution from a wide range of temperatures, creating a challenge for
obtaining unambiguous detailed constraints on the emitting plasma. 
For example, Craig and Brown (1976) showed that noise in the spectrum has a 
dramatic and unphysical impact on the plasma temperature distribution needed to recover that full spectrum,
including its noise.
To address this ill conditioning, many researchers have turned to a forward approach, where they choose some
specific low-order parametrizable form for 
the source distribution, sometimes using as few as three parameters to describe the shape of the
temperature dependence of the source distribution.
In this approach, it is assumed that a satisfactory fit to the data, interpolated through the noise, allows the chosen
parametrization to faithfully characterize the actual source plasma.
The advantage of this approach is that random and systematic uncertainties are not amplified by an ill-conditioned
inversion, the disadvantage is that ambiguity is introduced by the non-uniqueness in the form of the chosen parametrization.

When collisional ionization equilibrium (CIE) is assumed, the thermal X-ray spectrum is completely specified by the
differential emission measure (DEM) distribution $\DEM$ (Cox \& Tucker 1969; Raymond \& Smith 1977) 
%% Cox, D. P. \& Tucker, W. H. 1969, ApJ, 157, 1157
%% Raymond,  J. C. \& Smith, B. W. 1977, ApJS, 35, 419
over temperature $T$.
Thus it is usually accepted that analysis of an X-ray spectrum should involve inferring the DEM responsible for that spectrum,
and thus it is also common to parametrize the DEM as a means of achieving a fit to the data.
An implicit assumption of this approach is that an inferred DEM will then be physically meaningful for constraining theoretical models,
and the advantage of fitting the DEM is that it makes no additional assumptions about such models, beyond CIE.
However, the choice of the form of the parametrization introduces a kind of insidious model dependence, because of the ambiguity problem
alluded to above.
Theoretical models may attempt to fit particular forms of
the DEM when other quite different forms could suggest a rather different formation mechanism, yet still fit the data succesfully.
Hence, it may actually serve the modeler better to consider source parametrizations that are more conveniently tailored to the
physics actually being modeled.
Given that X-ray data often contains only a low degree of independent information about the source distribution, it is natural to attempt to
fit it with a low number of 
adjustable parameters, stressing the advantages of choosing such a parametrization to work closely with a model of particular interest, and
weakening the value of
a so-called ``model independent'' approach.
Since at the end of the exercise, if a source distribution that fits the data is achieved, there is no loss of generality regardless of
the parametrization scheme chosen.
It is only if the fitting efforts are frustrated that the guiding model would have proven to be ill suited for the task, a result that would 
come with its own lessons of interest.

This paper suggests one such alternative parametrization, tailored for
situations where the gas is thought to be heated impulsively up to a initial temperature, 
or range of initial temperatures, followed by transient cooling
through all lower temperatures.
The result is similar to models of cooling flows (e.g. Johnstone et al. 1992; Peterson et al. 2001; Mukai et al. 2003),
%% Johnstone, R. M., Fabian, A. C., Edge, A. C., \& Thomas, P. A. 1992, MNRAS, 255, 431.
%% Peterson, J. R., Paerels, F. B. S., Kaastra, J. S., Arnaud, M., Reiprich, T. H., Fabian, A. C., Mushotzky, R. F., Jernigan, J. G., \&
%% Sakelliou, I. 2001, A\&A, 365, 104.
%% Mukai, K., Kinkhabwalla, A., Peterson, J. R., Kahn, S. M., \& Paerels, F. 2003, ApJL, 586, L77.
except that the heating and cooling is stochastic and locally transient here, being steady only in a statistical sense when integrated over
the entire source region, and plasma can be injected over a distribution of initial temperatures.
The latter attribute allows greater flexibility in parametrizing the globally steady-state yet locally
impulsive heating distribution, by tracking, cumulatively over temperature $T$, the rate that heated plasma
is impulsively injected above each $T$.
The main contrast with a DEM approach is that while the DEM constrains an instantaneous snapshot of the source distribution, per temperature
bin $dT$,
the cumulative initial-temperature (CIT) distribution constrains the initial $T$ that the plasma reached in
its recent impulsively-heated and transient history.
Thus the CIT includes all the plasma heated above each $T$, so it is a cumulative
distribution over $T$, not a distribution over instantaneous $dT$ bins in a snapshot.
This takes the model one step farther from the data, but one step closer to the heating model, which may be considered a strength or
a weakness depending on the application.
To explore the contrasts in these parametrizations, this paper will focus on
one particularly simple yet 
fairly common type of forward-modeled DEM distribution, a two-temperature plasma.
That source model has proven effective in a wide variety of contexts, so the goal here is to understand how an alternative CIT parametrization
would function in those same contexts.

Applications where successful 2-$T$ fits have been applied to thermal X-ray spectra 
include RS CVn stars and active binaries (e.g., Swank et al. 1981),
%% Gehrels, N. \& Williams, E. D. 1993, ApJL, 418, L25.
hot-star winds (e.g., Zhekov et al. 2011),
%% Zhekov, S. A., Gagne, M., \& Skinner, S. L. 2011, ApJL, 727, L17. 
solarlike coronae (e.g., G{\"u}del et al. 2008),
%% Gudel, M., Skinner, S. L., Audard, M., Briggs, K. R., \& Cabrit, S. 2008, A\&A, 478, 797.
supernova remnants (e.g., McEntaffer \& Brantseg 2011; McEntaffer et al. 2013),
%% McEntaffer, R. L. \& Brantseg, T. 2011, ApJ, 730, 99.
and diffuse galactic emission (e.g., Kuntz et al. 2003; Kuntz \& Snowden 2008).
%$ Kuntz, K. D. \& Snowden, S. L. 2000, ApJ, 543, 195.
The fact that two-temperature fits are useful in these contexts leaves open the question of whether or not this should be taken
as evidence that the plasma is really being maintained continuously in two separate temperature regimes, or if this structure
reflects no physics beyond the arbitrary fitting procedure.
Thus it is important to investigate in general terms whether
impulsive heating and transient temperature structures could fit similar data, to within the experimental and theoretical uncertainties,
and that is a key goal of this paper.

Note that a two-$T$ DEM fit actually involves four parameters,
one a total emission-measure diagnostic (the sum of the two components), and
three shape parameters that affect the relative fluxes in the various lines and continuum energy bins.
The three shape parameters include the two $T$ chosen, and the ratio of the emission measures in the two components.
A CIT fit with four analogous parameters would include an overall total rate that particles are encountering the impulsive heating
(or passing through shocks), and three shape parameters including two initial $T$ for impulsive heating effects, 
and the ratio of the rates for those two initial $T$.
This might represent, for example, a standing forward/reverse shock pair with two different mass fluxes into the shocks.
Alternatively, a smoothly continuous CIT (albeit monotonically decreasing, as always in this model) 
could be used with three other shape parameters, or even fewer,
and might also mimic a two-$T$ fit in situations to be explored.
These possibilities are examined to answer the
fundamental questions, when can either approach be used to fit a spectrum, and how would the corresponding attributes
of such mutually possible fits be interpreted?
In particular, can we use the spectral shape to determine whether the gas is maintained at locally steady $T$, or impulsively
heated and transiently cooled?
In situations where two-$T$ DEM fits are already known to yield
satisfactory agreement with observations, can CIT fits work as well, and what is the physical significance?
%In situations where either approach achieves a fit, do we access more useful information
%by interpreting the spectral contraints in the light of some particular model to navigate this ambiguity?

The fundamental goal is to critically examine the concept of model independent X-ray fitting, in light of an alternative approach
of intentionally focusing on impulsive heating and seeking self-consistent evidence in favor of such a model.
As the conclusions are intended to apply in a qualitative way to a broad range of datasets, the considerable power of detailed
line modeling will be foregone in favor of more analytic and heuristic expressions designed to achieve greater flexibility and
generalizability.

%A related issue is, when are DEM fits as truly model independent as they are often intended to be,
%and when do they reflect modeling biases instead?
%We can question whether or not
%model independence should even be the goal, when the data may be more informative within 
%some specific and intentionally model-dependent context.

\section{Thermal X-ray Source Parametrizations}
%% kgg: insert remarks about the bremsstrahlung continuum

The two primary thermal X-ray diagnostics are the bremsstrahlung continuum, and the fluxes in the various lines present in the spectrum.
Let the spectral variable for the bremsstrahlung continuum be the scaled energy $\energy = E/k\To$, and let the lines be indexed by
their scaled temperature of peak emissivity 
$\ti = \Ti/\To$, where $E$ is the photon energy, $\Ti$ is the plasma temperature of peak contribution to the
line, and $\To$ is a fiducial temperature scale.
Here we choose $\To$ to be the temperature where the bremsstrahlung continuum radiative
cooling function $f_c(T)$ crosses the line radiative cooling function $f_l(T)$, so $f_c(\To) = f_l(\To)$.
The total radiative cooling function is $f_{rad}(T) = f_l(T) + f_c(T)$, and when multiplied by the
free electron density $n_e$ and the quantity of gas in some given temperature bin (often represented by the number of hydrogen atoms), 
gives the total rate of radiative energy generation in that bin.
Note the well-known result that in CIE with the further assumption that the dominant rates are collisional excitation and radiative
de-excitation, all these functions depend only on temperature $T$.
In order to manipulate
dimensionless order-unity quantities, in what follows
the temperature $T$ will also be scaled to $\To$ via $\temp = T/\To$, joining the photon energy $\energy$ and the characteristic line
temperatures $\ti$ as the fundamental variables of interest here.

\subsection{General expressions}
%   source times kernel, kernel normalization

In this notation, the bremsstrahlung continuum energy flux per wavelength takes the general form
\beq
\label{fcgeneral}
F_c(\energy) \ = \ \int_0^\infty d\temp \ S(\temp) K_c(\temp,\energy) \ ,
\eeq
where $S(\temp)$ is the source rate of radiative energy in the $d\temp$ bin, and $K_c(\energy,\temp)$ gives the probability
that energy radiated from the $d\temp$ bin appears in the bremsstrahlung continuum within $d\energy$ of the photon energy $\energy$.  
Similarly, the flux in a given line $i$, indexed by its characteristic temperature of peak emissivity $\ti$, is
\beq
\label{figeneral}
F_i(\ti) \ = \ \int_0^\infty d\temp S(\temp) K_i(\temp,\ti) \ ,
\eeq
where $K_i(\temp,\ti)$ gives the probability that energy radiated from the $d\temp$ bin appears in line $i$, rather than in the other
lines or the continuum.
The index $i$ by itself identifies the line, but $\ti$ is also included explicitly here because the approach below will be to choose
a generic form for $K_i(\temp,\ti)$, and regard $\ti$ as a (hypothetical) continuous variable.

\subsubsection{The standard differential emission measure distribution (DEM)}
%   source is EM times cooling rate

In CIE, the standard way to write the source rate of radiated energy, $S(\temp)$, is to multiply a $T$-dependent radiative cooling
function, $\frad(\temp)$, by a differential emission measure (DEM), $\DEM(\temp)$, where
\beq
\DEM(\temp) \ = \ n_e n_H \frac{dV}{dT}
\eeq
and $\frad(\temp)$ can be tabulated via detailed atomic calculations (Sutherland \& Dopita 1993; though potential modifications
may be required, De Rijcke et al. 2013), 
%% Sutherland R. S., Dopita M. A. ApJS 1993;88:253.
%% De Rijcke, S., Schroyen, J., Vandenbroucke, B., Jachowicz, N., Decroos, J., Cloet-Osselaer, A., Koleva, M. 2013, MNRAS, 433, 3005
or approximated more
heuristically as is done here.
If we further break up $\frad(\temp)$ into its dominant contributions from lines and the bremsstrahlung continuum as
\beq
\frad(\temp) \ = \ f_l(\temp) \ + \ f_c(\temp) \ ,
\eeq
we then have
\beq
\label{sdem}
S(\temp) \ = \ \DEM(\temp) \To [f_l(\temp) \ + \ f_c(\temp)] \ .
\eeq
Hence the conventional approach for dealing with eqs. (\ref{fcgeneral}) and (\ref{figeneral}) is to parametrize $\DEM(\temp)$
and use the atomic physics to predict the observed rate of line and continuum radiation.
However, the $\DEM(\temp)$ distribution does not by itself convey any immediate information about the processes responsible
for the hot gas, that physics has to be provided by subsequent modeling efforts that are presumably capable of reproducing a similar
DEM distribution.
However, when the DEM is parametrized to obtain a fit, biases are introduced by the form of the parametrization,
so although it is normally regarded as ``model independent'' to fit the DEM, in practice this is not necessarily the case.
As such, it behooves us to consider other forms of parametrization that may, in some situations, convey more directly physically 
relevant constraints.

\subsubsection{The cumulative initial-temperature (CIT) distribution for impulsive heating}
%with complete cooling}
%   source is p(T) times the radiative efficiency

One such alternative parametrization involves imagining that the gas is being impulsively heated at some prescribed rate over a
distribution of initial temperatures, here rescaled similarly as a unitless $\temp$, and then cools transiently through all lower temperatures.
This approach is reminiscent of models of steady-state cooling flows (e.g., Peterson \& Fabian 2006),
%% Peterson, J. R. \& Fabian, A. C. 2006, PhR, 427, 1
except that here the heating is only steady in a statistical sense, so may be stochastically distributed in space, perhaps by a range of shock
strengths and obliquities.
Since the gas is assumed to cool through all lower temperatures, all that needs to be specified is the 
cumulative distribution over initial temperature (CIT),
$\CIT(\temp)$, and the resulting radiative emission is given by eqs. (\ref{fcgeneral}) and (\ref{figeneral}) in concert with
\beq
\label{scpt}
S(\temp) \ = \ \CIT(\temp) \alpha k \To S_o \frac{[f_l(\temp) \ + \ f_c(\temp)]}{[f_l(\temp) \ + \ f_c(\temp) \ + \ f_o(\temp)]} \ .
\eeq
The CIT distribution is cumulative in the sense that the gas cooling through the $d\temp$ bin in the integral over the sources includes all
the gas heated above $\temp$.
Here $\alpha k$ is the energy released per particle per $dT$ that the gas cools through,
$S_o$ is the total rate that particles are introduced to the impulsive
heating (say, the rate they pass through shocks), $\CIT(\temp)$ is the distribution over initial temperatures normalized by
$\CIT(0) = 1$ (since the probability that any shock or impulsive heating even yields an initial temperature 
above $\temp = 0$ is unity), and $f_o(\temp)$ accounts for the nonradiative cooling rate.
As such, eq. (\ref{scpt}) involves a total energy release rate multiplied by a radiative branching ratio, where $f_o(\temp)$ includes
the relative rate of internal energy dissipation from adiabatic expansion, conduction, and mixing of hot and warm gases.
The difficulty in constraining this term results in the model dependence of this approach, but this added complexity is
mitigated in situations where efficient radiative
cooling may be assumed.
The special case of a cooling flow has a flat $\CIT(\temp) = 1$ up to the $\temp$ of the environment feeding in the
cooling gas, which is functionally equivalent to an impulsive heating mechanism that always produces the same initial $\temp.$

Note that the interpretations of the kernel functions $K_c(\temp,\energy)$ and $K_i(\temp,\ti)$ are slightly altered in the CIT approach.
Here $K_c(\temp,\energy)$ is the fraction of the radiated energy that appears 
in the bremsstrahlung continuum within $d\energy$ of
$\energy$, as the gas cools through $\temp$.
Similarly,  $K_i(\temp,\ti)$ is the  fraction of the radiated energy that appears
in line $i$, as the gas cools through $\temp.$
The kernels are the same as in the DEM picture, where here the DEM is only steady in a globally stochastic and locally transient sense.

\subsubsection{The conversion between the CIT and the DEM}
%  equivalence

The consistent form of eqs. (\ref{sdem}) and (\ref{scpt}) allows for a direct conversion between the two parametrization schemes, given by
\beq
\label{conversion}
\CIT(\temp) \ = \ \DEM(\temp) \frac{[f_l(\temp) \ + \ f_c(\temp) \ + \ f_o(\temp)]}{\alpha k S_o} \ .
\eeq
Given this expression, the two parametrizations are formally equivalent, but they will differ in practice because the total cooling
function is subsumed into the $\CIT(\temp)$ parametrization, making it fundamentally a parametrization of the heating process independently
of the cooling, a feature that the usual DEM parameterization does not possess.
Also, the use of the CIT parametrization connects the total radiative emission to the normalizing parameter $S_o$, which is intended to
convey direct information about the total rate that the gas is encountering the heating processes.
A DEM-type parametrization, on the other hand, conveys an overall scale that is not as easily interpreted.
Neither of these advantages to the CIT approach are applicable unless the overall impulsive heating model applies, and the reliance on
information about $f_o(\temp)$ implies that the CIT approach does not inherit all the model independence of the DEM approach, despite
this formal conversion.

%\subsection{Simplifications for efficient radiative cooling}
% how CIT simplifies if all cooling is radiative

\subsection{Discussion}
%     (a) loss of model independence may be a feature, not a bub
%     (b) C(T is especially simple if F_o = 0, for then it not only has the same model independence as the DEM, it
%            introduces no artificial dependence on the radiatvie cooling function
%     (c)C(T must be monotonic (gives a self-consistency check)
%     (d)Even if sophisticated DEM methods are used, knowledge of the heating still requires comparing the results
%      to some simpler parametrization, so why not just find the simple parametrization that best fits the data
%      in the first place?

As mentioned above, the CIT approach involves additional model dependence, because of the need to specify $f_o(\temp)$, and also
because it receives its advantages only when the heating is impulsive, especially for a distribution of shock strengths.
This limitation is mitigated by the conceptual advantages, when applicable,
because a physically motivated parametrization yields more direct conclusions about the heating process.
Since ill conditioning of the spectral inversion induces ambiguity in the source model, we have only a few adjustable parameters
available for extracting the constraints in the data, so it may well behoove us to choose those parameters judiciously.
Such judicious choices always involves some form of model dependence-- for example, popular multitemperature DEM fits suggest an expectation
that the plasma can be held continuously at just a few characteristic temperatures, so is of less practical interest
if the gas actually cools through
all lower temperatures.

Moreover, the model dependence in the CIT approach is removed if we can confidently assert efficient radiative cooling and
thus take $f_o(\temp) = 0$.
In that situation, the CIT approach is mathematically equivalent to any DEM approach, yet involves parametrizations
that lack the implicit
dependence on the radiative cooling function that appears in the DEM, as seen in eq. (\ref{conversion}).
In other words, there will always be reduced DEM at temperatures where the cooling is most efficient, simply because it is
difficult to maintain as much plasma at those temperatures, but fitting that reduction
does not constrain a heating mechanism that is presumably independent of the radiative cooling function.
In such a case, fitting the CIT constrains the heating more directly, such that none of the flexibility offered by the fit
parameters is being deployed to simply track variations in the cooling function.

Even more importantly, if the heating is impulsive, and is followed by efficient radiative cooling through all lower $T$, then
the inferred $\CIT(\temp)$ function must be {\it monotonically decreasing} with $\temp$, because the CIT distribution is cumulative.
Thus the spectral fit gives an immediate internal consistency check, where if lower-temperature lines do not meet the
minimum flux requirements implied by the presence of the higher-temperature lines, then it must not hold that the gas
cools through all the lower temperatures.
Also, if a DEM approach is used and a fit to $\DEM(\temp)$ is obtained that maps into a monotonically decreasing 
$\CIT(\temp)$ via eq. (\ref{conversion}), this suggests the possibility that the gas is not being maintained continuously at the
temperatures in the DEM, but may instead be transiently cooling.
Of course, it may also be that there is simply more gas being continuously maintained at lower temperature, but at
least the CIT formalism provides a consistency check on the assumed physics, whereas DEM fits allow any possibility and so provide
no such consistency checks.
Again, it is a case where some model dependence in the spectral fitting may actually help guide appropriate modeling choices.

Finally, a benefit of the CIT approach is that the parameters obtained in the fit actually provide direct constraints on the heating,
if it is indeed impulsive and radiatively cooled.
The $\CIT(\temp)$ function that is produced gives the cumulative rate that gas is heated above each $\temp$ times the radiative
branching fraction, so if the radiative branching is near unity, then features in the
fit map directly into features in the heating mechanism.
If the radiative branching cannot be assumed to be unity due to a possible role for nonradiative cooling, then nonmonotonic features 
in a CIT fit can be attributed to variations in the nonradiative cooling, such as if lower temperature gas was mixing with even cooler
gas and creating a defecit in low-temperature line emission.
Such features are potentially physically interesting, unlike features in the DEM that exercise the fit parameters yet stem simply from
variations in the radiative cooling curve.

\section{Approximate Expressions for Radiative Cooling and the Kernel Functions}

Since the goal of this paper is a general broad-brush analysis, the line and continuum radiative cooling functions
$f_l(\temp)$ and $f_c(\temp)$, and the line and continuum emission kernels $K_i(\temp,\ti)$ and
$K_c(\temp,\energy)$ will be approximated in simple analytical forms.
These approximations are intended to convey similar qualitative behavior as including detailed kernels for the
known atomic physics, yet without complicating the resulting expressions with corrections for Gaunt factors or specific atomic
level structures that require detailed atomic models to make progress.
Certainly such models are essential for extracting diagnostic information from real spectra, but the purpose here is simply
to understand the relative diagnostic potential of the DEM and CIT parametrizations, so a simpler and more analytic treatment is indicated.

\subsection{Approximating the radiative cooling functions}

The line radiative cooling function $f_l(\temp)$ is a detailed curve 
with several bumps and features, but in broad-brush is generally
decreasing with $T$ in the X-ray regime owing to the stripping of bound electrons.
A rough qualitative fit can be obtained using
\beq
f_l(\temp) \ \cong \ A_o \temp^{-1/2} \ ,
\eeq
where $A_o$ is a constant that is not of interest in this paper because it is the relative shape
of the spectra, not overall intensity, that is of interest for the comparisons made here.
The bremsstrahlung continuum cooling function is smoother (Karzas \& Latter 1961; Kellogg, Baldwin, \& Koch 1975),
%% Karzas, W. J. \& Latter, R. 1961, ApJS, 6, 167
%% Kellogg, E.; Baldwin, J. R.; Koch, D. 1975, ApJ, 199, 299
but still includes complicated Gaunt-factor quantum corrections.
Taking a unit Gaunt factor for simplicity yields
\beq
f_c(\temp) \ \cong \ A_o \temp^{1/2} \ ,
\eeq
where note $f_c(1) = f_l(1)$ as per the defining convention for the unitless temperature $\temp$.
These approximations yield generally correct qualitative behavior, but quantitatively reliable
analyses must instead include the detailed atomic physics.

\subsection{Approximating the kernel functions}

The bremsstrahlung continuum kernel function with unit Gaunt factor is
\beq
\label{kcgaunt}
K_c(\temp,\energy) \ \cong \ \frac{e^{-\energy/\temp}}{\temp} \frac{f_c(\temp)}{[f_c(\temp)+f_l(\temp)]} \ ,
\eeq
normalized such that
\beq
\int_0^\infty d\energy \ K_c(\temp,\energy) \ = \ \frac{f_c(\temp)}{[f_c(\temp)+f_l(\temp)]} \ ,
\eeq
as is appropriate given its interpretation as the probability per $d\energy$ that energy radiated by gas at $\temp$ will appear
at photon energy within $d\energy$ of $\energy$.
Given the above approximations for the radiative cooling functions, this results in simply
\beq
\label{kcapprox}
K_c(\temp,\energy) \ \cong \ \frac{e^{-\energy/\temp}}{(1+\temp)} \ ,
\eeq
which then gives
\beq
\label{fc}
F_c(\energy) \ \cong \ \int_o^\infty d\temp \ \DEM(\temp) \frac{e^{-\energy/\temp}}{\sqrt{\temp}} \ = \ 
\int_o^\infty d\temp \ \CIT(\temp) \frac{e^{-\energy/\temp}}{(1+\temp)}
\eeq
in the two parametrizations considered here.

A heuristic approximation of the line kernel must reflect the possibility that some lines have a wider temperature response
in their emissivity functions than do others.
Most lines have a full-width-half-maximum (FWHM) in
their temperature response of a factor of 2--3 in temperature, and by
convention the line emissivity function should peak at $\temp = \ti$, so a reasonable approximation is offered by
\beq
\label{kiapprox}
K_i(\temp,\ti) \ \cong \ A_i \Deltati \frac{n^{n-1}}{\Gamma(n-1) \ti} \left ( \frac{\ti}{\temp} \right )^n e^{-n\ti/\temp} \ ,
\eeq
where $A_i$ is the abundance relative to some standard,
and $\Gamma(x)$ is the standard Gamma function, so equals $(n-2)!$ when $n$ is an integer.
Here $n$ is an adjustable parameter that measures the width of the kernel function in temperature space, where $n = 5$
corresponds to a FWHM of about a factor of 2 in $\temp$, and $n = 12$ gives a FWHM of about a factor of 3, which are
both fairly typical for X-ray lines.
In what follows, it will be seen that the temperature diagnostics afforded by lines with $n \cong 5$ is substantially
higher than those with $n \cong 12$, hence the need for a flexible line kernel form.
The normalization of $K_i(\temp,\ti)$ is such that
\beq
\int_0^\infty d\temp \ K_i(\temp,\ti) \ = \ A_i \Deltati
\eeq
where for simplicity in this paper $A_i = 1$, though in general it could be varied to improve the model fit,
and the meaning of $\Deltati$ will be discussed next.

\subsection{The temperature equivalent width of a line}
 
It is natural to expect the flux in any line to be proportional to some type of quantitative line strength, so the diagnostic
potential in each line is not so much in the flux, but in the factor by which that flux deviates from this expected line strength.
One way to characterize the line strength is to consider its peak emissivity, occuring for $\temp = \ti$, and indeed this is more
or less the approach taken by isothermal line diagnostics that ask how much emission measure, at a given temperature, would be
needed to produce the observed flux in the line in question.
However, assuming the source distribution is a relatively smooth function of temperature, rather than isothermal, an alternative
approach is to assume the DEM is flat over the temperatures of significant line response, and take the line strength to be the
integral of $K_i(\temp,\ti)$ over $\temp$, which here is the parameter $\Deltati$ in units of the fiducial temperature $\To$.
This quantity has the meaning of an ``equivalent temperature width,'' in the sense that it gives the width in $\temp$ that would
produce the same line flux as the actual emissivity function, if $K_i(\temp,\ti)$ were unity over that $\temp$ regime, and
zero at all other $\temp$.
Physically, the expected flux in any line is proportional to the $\Deltati$ for that line, if the total emission rate
is flat over the line-forming
temperature domain.
Using this quantity to characterize the line strength represents a break from the standard method of analyzing 
the emission measure all at one temperature that would be needed to yield the line flux at that temperature, because that
approach emphasizes the temperature of peak emissivity for each line, when extracting constraints from the data.
Hence the new approach is more physically appropriate for smooth source distributions rather than isothermal ones, and smooth
distributions seem generally easier to obtain in models that follow some physically realizable mechanism.

\subsection{Normalized line flux (NLF) distribution over peak-temperature space}

Since we expect line fluxes to be proportional to their temperature equivalent widths for a flat DEM, deviations from this expectation
is what actually constrains the DEM, not the observed line fluxes themselves.
Hence it makes sense to normalize each line flux $F_i(\ti)$ by its
temperature equivalent width at standard abundance $\Deltati$, and its relative abundance $A_i$ to that standard,
thereby producing a normalized line flux (NLF) distribution
\beq
\fline(\ti) \ = \ \frac{F_i(\ti)}{k A_i \Deltati} \ ,
\eeq
where $k$ is the Boltzmann constant that puts $\Deltati$ in energy units,
and in this paper $A_i = 1$ for simplicity, though in general the need to yield a smooth and consistent $\fline{\ti}$ can serve as a guide
to the actual abundances.
Note that the NLF distribution, $\fline{\ti}$, is a kind of
line spectrum in the space of temperature of maximum line emissivity, $\ti$, 
not it photon energy space.
It can also be thought of as a way to connect each line flux with a concept of an average CIT value at the each line formation temperature $\ti$,
because if we define
\beq
\left < \psi \right >_i \ = \ \frac{\int_0^\infty d\temp \ K_i(\temp,\ti) \psi(\temp)}{\int_0^\infty d\temp \ K_i(\temp,\ti)} \ ,
\eeq
then it follows that
\beq
\fline(\ti) \ = \ \frac{\alpha}{k} \left < \psi \right >_i  \ ,
\eeq
which reveals the fundamental constraint on the source distribution that is afforded by the observationally determined NLF distribution.

Note that the NLF distribution is regarded as a smooth curve in this approach, though in actual
fact it is sparsely populated by lines at specific $\ti$.
Its general shape provides
the primary constaints on the CIT shape, and via eq. (\ref{conversion}) the DEM shape also, that can be extracted from lines.
In the general form of source times kernel used above, this motivates defining the normalized line kernel by
\beq
\kline(\temp,\ti) \ = \ \frac{K_i(\temp,\ti)}{A_i \Deltati} \ ,
\eeq
which then gives
\beq
\fline(\ti) \ = \ \int_0^\infty d\temp S(\temp) \kline(\temp,\ti) \ .
\eeq
In the heuristic approximations used here, we thus have
\beq
\kline(\temp,\ti) \ = \ \frac{n^{n-1}}{\Gamma(n-1) \ti} \left ( \frac{\ti}{\temp} \right )^n  e^{-n \ti/\temp} \ ,
\eeq
which is normalized by $\int_o^\infty d\temp \ \kline(\temp,\ti) \ = \ 1$, and finally
\beq
\fline(\ti) \ \cong \  \int_0^\infty d\temp S(\temp) \frac{n^{n-1}}{\Gamma(n-1) \ti} 
\left ( \frac{\ti}{\temp} \right )^n  e^{-n \ti/\temp} \ .
\eeq

The two separate source parameterizations in eqs. (\ref{sdem}) and (\ref{scpt}) then combine with eq. (\ref{fline}) to yield
the fundamental line diagnostics analyzed here, given by
\begin{align}
\begin{split}
\label{fline}
\fline(\ti) \ &\cong   \int_0^\infty d\temp \ \DEM(\temp) \left ( \sqrt{\temp} + \frac{1}{\sqrt{\temp}} \right )
\frac{n^{n-1}}{\Gamma(n-1) \ti}
\left ( \frac{\ti}{\temp} \right )^n  e^{-n \ti/\temp} \\ 
&=  
\int_0^\infty d\temp \ \CIT(\temp) 
\frac{(\sqrt{\temp} + 1/\sqrt{\temp})}{(\sqrt{\temp} + 1/\sqrt{\temp} + f_o)}
\ \frac{n^{n-1}}{\Gamma(n-1) \ti}
\left ( \frac{\ti}{\temp} \right )^n  e^{-n \ti/\temp} \ ,
\end{split}
\end{align}
where again $\ti$ is regarded as a hypothetical continuous input variable, though of
course any real line list will present only specific peak temperatures.
The NLF distribution $\fline{\ti}$ presents a kind of flux spectrum
in plasma $T$ space, and constrains the shape of the source distribution moreso than unnormalized line fluxes, because the NLF
allows us to see the deviations in the line fluxes from what a globally flat DEM distribution would give.

\subsection{Monotonicity of both the bremsstrahlung continuum and the NLF distribution in CIT}

In the case of highly efficient radiative cooling, eq.(\ref{scpt}) shows that the CIT source distribution $S(\temp)$ inherits
the monotonically decreasing nature of $\CIT(\temp)$, as required by the cumulative distribution.
It immediately follows that both $F_c(\energy)$ and $F_i(\ti)$ must also be monotonically decreasing for the approximations here.
The bremsstrahlung continuum $F_c(\energy)$ must be monotonically decreasing with $\energy$ for any source distribution when
unit Gaunt factor is used,  because
eq. (\ref{kcgaunt}) approximates the kernel as monotonically decreasing with $\energy$, so eq. (\ref{fcgeneral})
shows that $F_c(\energy)$ will be also.
Thus monotonicity in the bremsstrahlung continuum is a basic requirement that does not constrain the sources, but is
useful for seeing the onset of absorption at low $\energy$.

The line kernel, on the other hand, is not monotonically decreasing in $\ti$, so this allows $F_i(\ti)$ to provide
useful constraints via its monotonicity or nonmonotonicity, 
because $F_i(\ti)$ does monotonically decrease with $\ti$ in CIT-type parametrizations but not necessarily DEM-types, 
for the approximations used here with efficient radiative cooling.
This follows from a theorem that if
\beq
\label{theorem}
F(x) \ = \ \int_0^\infty dy \ \frac{G(y)}{y} P \left ( \frac{x}{y} \right ) \ ,
\eeq
and if $G(y)$ is monotonically decreasing, then $F(x)$ is also.
This can be seen  by making the replacement $z = x/y$, since then
\beq
F(x) \ = \ \int_0^\infty dz \ \frac{G(x/z)}{z} P(z) \ ,
\eeq
and that is obviously monotonically decreasing with $x$ because $G(x/z)$ is.
This theorem implies that the NLF distribution $F_i(\ti)$ decreases monotonically with $\ti$ because $\psi(\temp)$ decreases
monotonically with $\temp$, as can be seen by noting the close connection between eq. (\ref{theorem}) and
the fact that for complete radiative cooling, eqs. (\ref{figeneral}), (\ref{scpt}), and (\ref{kiapprox}) give
\beq
F_i(\ti) \ = \ \frac{n^{n-1}}{\Gamma(n-1)} \alpha k \To S_o A_i \Delta T_i \int_0^\infty d\temp \ 
\frac{\psi(\temp)}{\temp} \left ( \frac{\ti}{\temp} \right )^{n-1} e^{-n\ti/\temp} \ ,
\eeq
so we need merely associate $\psi$ with $G$ and use the appropriate $P$ to apply the theorem.

This provides an easy operational method to know when a radiatively cooled CIT-type model, such as a cooling flow or any type of 
stochastic impulsive heating followed by complete radiative cooling, might be
appropriate.
Such a model is worth 
consideration whenever the NLF distribution is monotonically decreasing, once
foreground absorption (not included in these expressions) is removed appropriately, or at least sufficiently 
that the bremsstrahlung continuum decreases
monotonically as well.
Of course, nonradiative cooling processes at low $T$ could reduce the
monotonicity of $F_i(\ti)$ within the context of impulsively heated sources, so that should be taken into consideration
when monotonicity in the NLF is violated at low $T$.

\subsection{Connections with the Laplace transform}

The above approximate kernels allow $F_c(\energy)$ and $\fline(\ti)$ to be expressed as Laplace transforms
of the appropriately translated source terms.
Taking $\laplace[]$ to denote the Laplace transform from $x = 1/\temp$ to either $\energy$ (for the bremsstrahlung continuum)
or to $\ti$ (for line fluxes), so
\beq
\laplace \left [h(x) \right ] \ = \ \int_0^\infty dx \ h(x) e^{-x \energy} \ ,
\eeq
eqs. (\ref{fc}) and (\ref{fline}) yield
\beq
F_c(\energy) \ = \ \laplace \left [\frac{\DEM(1/x)}{x^{3/2}} \right ] \ = \  \laplace \left [\frac{\CIT(1/x)}{x(1+x)} \right ]
\eeq
and
\beq
\fline(\ti) \ = \  \frac{\ti^{n-1}}{\sqrt{n} \Gamma(n-1)} \laplace \left [x^{n-3/2}\left (1 + \frac{n}{x}\right )\phi \left (
\frac{n}{x} \right ) \right ] \ = \   \frac{\ti^{n-1}}{\Gamma(n-1)} \laplace \left [x^{n-2} \psi \left (\frac{n}{x} \right ) \right ] \ .
\eeq

This formal equivalence to Laplace transforms
is convenient, as it brings in a significant body of knowledge associated with those common transforms, and offers
insights into the degree of ill conditioning in the inverse problem.
In particular, the correspondence allows us to
know that smooth spectra modeled with these approximate kernels are uniquely invertible, such that the source
distributions could in principle be determined by the continuum spectrum via
\beq
\DEM(\temp) \ = \ \frac{1}{t^{3/2}} \laplace^{-1} \left [ F_c(\energy) \right ]
\eeq
and
\beq
\CIT(\temp) \ = \ \frac{(1+t)}{t^2} \laplace^{-1} \left [ F_c(\energy) \right ] \ ,
\eeq
where the Laplace transforms map from $\energy$ to a dummy variable $x$, which is then replaced by $x = 1/\temp$.
The source distributions could also be found from the line-flux spectrum via
\beq
\DEM(\temp) \ = \ \Gamma(n-1) \frac{\ti^{n-2}}{n^{n-2} (\sqrt{t} + 1/\sqrt{t})} \laplace^{-1} \left [\frac{\fline(\ti)}{\ti^{n-1}} \right ]
\eeq
and
\beq
\CIT(\temp) \ = \ \Gamma(n-1) \frac{\ti^{n-2}}{n^{n-2}} \laplace^{-1} \left [\frac{\fline(\ti)}{\ti^{n-1}} \right ] \ ,
\eeq
where this time the Laplace transforms map from $\ti$ to a dummy variable $x$ which is then replaced by $x = n/\temp$.
However, these formal inversions are misleading, because uncertainties
in the observations, and systematic errors in the approximate kernels, are amplified when inverting back to the source distribution, 
to the extent that the uniqueness of the inversions comes at the cost of results that would likely not yield consistency between
the line and continuum spectra, nor with any physically meaningful source distribution (Craig \& Brown 1976).
%%ref Craig, I. J. D. \& Brown, J. C. 1976, Nature, 264, 340.
Thus the problem is normally solved using a low-order parametrization of the source distribution, and seeking only an approximate fit
to the data, hence trading ill-conditioning for ambiguity in the set of parametrizations that might be deemed adequate.
In the presence of uncertainties at roughly the 20\% level, these ambiguities are significant enough
to permit widely different fit parametrizations, as
will be seen shortly.

\section{Fit Comparisons Between DEM and CIT Approaches}
%  say someting about ambiguity and the different information that different parametrized fits tell us

A common approach to thermal X-ray spectral fitting is to use a sophisticated atomic database in concert with a quite rudimentary
source distribution, such as an isothermal or two-temperature plasma.
This represents a kind of disconnect in precision, because it is unlikely that isothermal or two-temperature fits are precise representations
of the actual source plasma, yet the atomic models may be fairly detailed and accurate.
The goal in this section is to examine what insights can be obtained about the nature of such simplistic DEM and CIT type fits,
from a more approximate and heuristic, yet flexible and transparent, perspective.
As the results here are not quantitatively reliable, the intention is not to replace detailed modeling, but rather to learn something about
what types of detailed fit information is necessary to distinguish various types of models, particularly models where the gas is maintained
at fixed temperature (DEM type), and models where the gas is impulsively heated and then transiently cooled (CIT).
When can either approach be used, or when is one or the other going to provide greater insight?

\subsection{One-steady-$T$ versus one-initial-$T$ fits}

A complete analysis of the precision with which one can potentially diagnose truly isothermal plasma is given
in Judge, Hubeny, \& Brown (1997), here the focus is on the more general question of how well the DEM can be constrained
when an isothermal DEM provides a reasonable fit, yet it is not known that the plasma is actually isothermal at all.
Since an isothermal plasma at temperature $\tow$ provides the simplest possible DEM fit, it is natural to take this as
a kind of standard benchmark.
The simplest CIT fit involves impulsive heating to a single initial
temperature $\tinitial$,
followed by cooling through all lower temperatures, so this is a natural comparison to draw when asking how
easily these two general models can be distinguished.

Applying the above kernel approximations, Figure 1
%% figure 1  compare one-T and one-shock
\begin{figure}[h]
\begin{center}
\plotfiddle{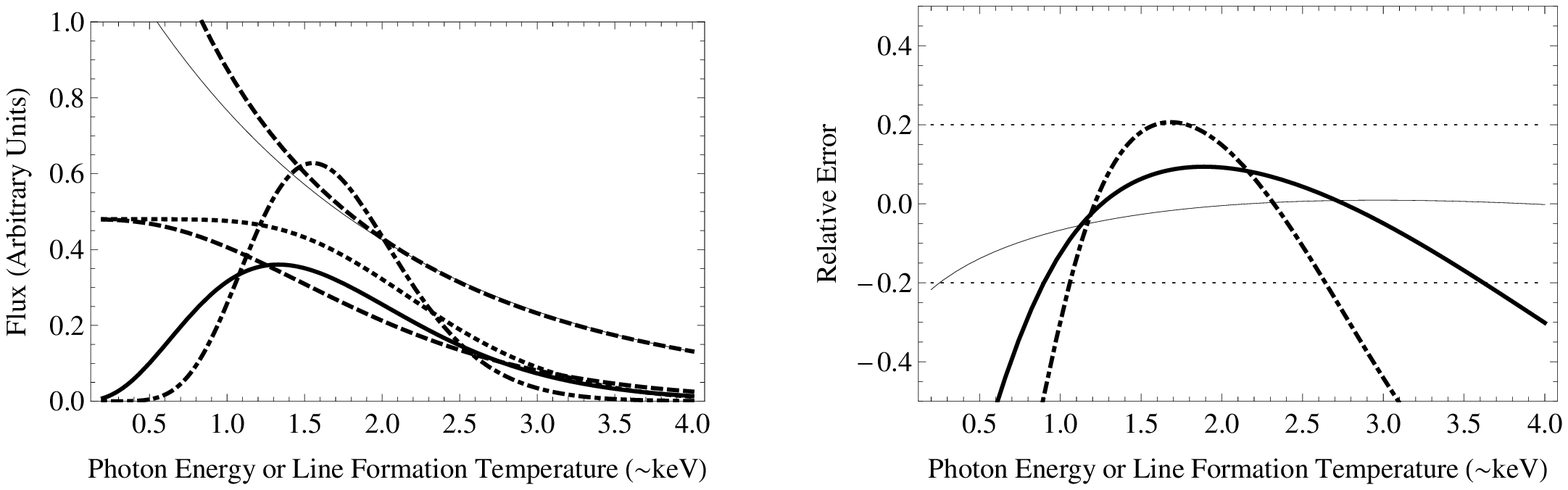}{2.6in}{-0.}{440.}{200.}{0}{0}
%\plotfiddle{PSFILE}{VSIZE}{ROTANG}{HSCALE}{VSCALE}{HTRANS}{VTRANS}
%\includegraphics[width=6.5in]{shockanglefig1.pdf}
\caption{
Left panel: The approximate bremsstrahlung continuum and heuristic
NLF distributions in arbitrary units, as a function of $\energy$ and $\ti$ 
respectively, comparing an isothermal DEM model with $\temp = 1.7$ and a single-initial-$T$ CIT model
with $\tinitial = 2.6$.
The isothermal bremsstrahlung continuum is the thin solid curve, the single-initial-$T$ bremsstrahlung
continuum is the dashed curve that follows along it,
and the NLF distributions for lines with emissivity functions with FWHM of a factor of 3 are shown with
the thick solid and dashed curves, for the isothermal and single-initial-$T$ models respectively.
Also shown are the NLF distributions for lines with emissivity functions with FWHM of a factor of 2, 
where the dot-dashed curve is for the isothermal DEM, and the dotted curve is for the single-initial-$T$ 
CIT model.  Right panel: The same curves, except showing the comparison of the difference divided by
the average for DEM and CIT models, where the thin curve is the bremsstrahlung continuum, the
thick curve is the NLF for lines with emissivity FWHM of a factor of 3 in $T$, and the dot-dashed
curve is for lines with FWHM of a factor of 2.
The dotted horizontal lines show the limits of 20\% uncertainty, intended as
a schematic estimate of reasonable experimental and theoretical errors.
}
\end{center}
\end{figure}
compares 
the relative shapes of the bremsstrahlung continuum for an isothermal model with $\tow = 1.7$ and
a one-initial-$T$ model with $\tinitial = 2.6$,
and the shapes of the NLF distribution for those two cases as well,
where note that if $\To \cong 1$ keV, then $\energy$ and $\ti$
are roughly in keV.
These roughly represent the best mutual fits that can be obtained with either isothermal or one-initial-$T$ parametrizations, but
the figure shows that the mutual fits are generally not good, except in the bremsstrahlung continuum at $\energy$ above about
3/4 of $\tow$, where the agreement is generally to within about 10\%.
The normalized line-flux distributions are given for two different line emissivity widths, corresponding to FWHM of a factor of
2 in $\temp$ space, and a FWHM of a factor of 3, and generally produce very poor agreement, especially for $\ti < \tow$, and the
lines with broader emissivity domains also produce a lot more flux at high $\ti$ in the one-initial-$T$ model than in the single-$T$ model.
The former problem is because the one-initial-$T$ model includes a range of cooler gas, and the latter problem is because the
initial $\tinitial$ in the CIT is chosen higher than the single $T$ in the DEM.
Although this is just a single example, it generally holds that one-initial-$T$ models are easily distinguished from single-$T$ models.
However, as these models have only one degree of freedom in describing the spectral shape, it is unlikely that either would fit a 
real dataset.

\subsection{Two-steady-$T$ versus one-initial-$T$ fits}
% Laplace transform surprise, relevance to 2-T fits by one characteristic shock strength

%It is often true that data is not well fit by an isothermal plasma, expressly because there is a need for cooler-temperature emission stemming
%from a wider array of lines.
When single-parameter models fail, 
the next simplest multi-temperature fit invokes two discrete $\temp$, with a separate emission measure (EM) for each.
This increases the number of spectral shape parameters from one to three, the two $\temp$ and the ratio of their EM, so one may
expect a significant increase in fitting flexibility when using a two-$T$ DEM form.
When a fit is achieved that way, it involves cooler plasma than the higher $\temp$, so a one-initial-$T$ CIT style parametrization may agree
better with a two-$T$ DEM than a one-$T$ DEM, in some parameter regime that can be explored using the heuristic expressions here.
Indeed, we shall soon see that in some cases, the bremsstrahlung continuum
is {\it virtually indistinguishable} in a two-$T$ DEM and a one-initial-$T$ CIT model, 
and the NLF may also appear similar, except at $\ti$ lower than roughly 3/4 of the
cooler $\temp$ in the two-$T$ fit.

Just such a situation is depicted in Figure 2,
%% figure 2, two-T vs. one-shock, showing Laplace surprise
\begin{figure}[h]
\begin{center}
\plotfiddle{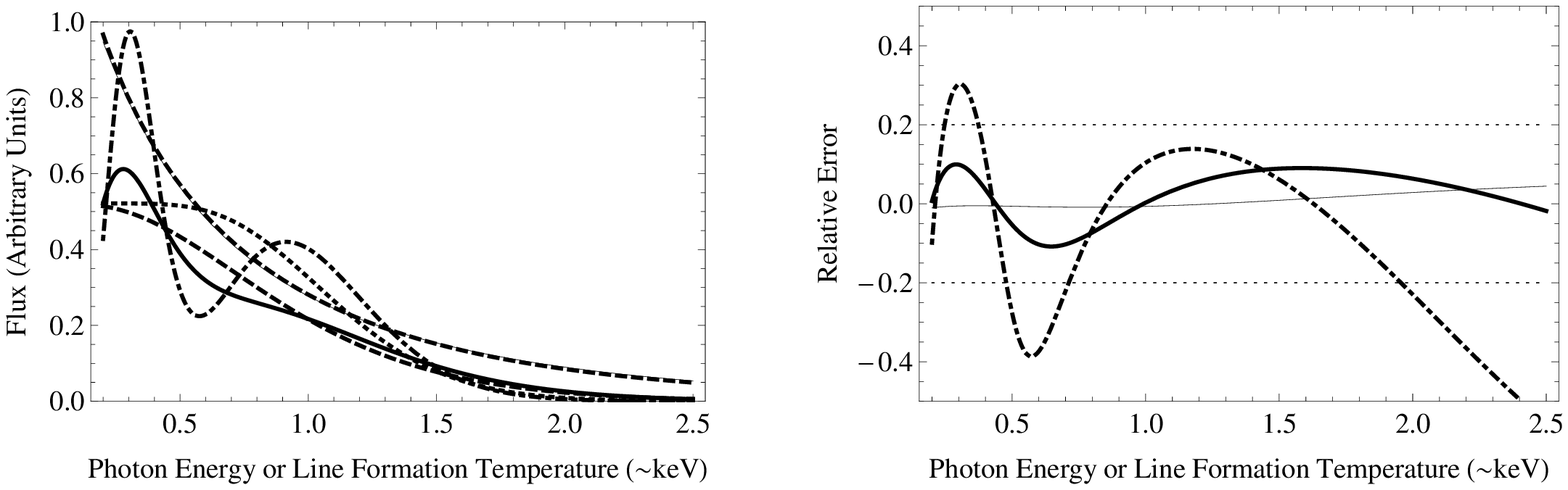}{2.6in}{-0.}{440.}{200.}{0}{0}
%\plotfiddle{PSFILE}{VSIZE}{ROTANG}{HSCALE}{VSCALE}{HTRANS}{VTRANS}
%\includegraphics[width=6.5in]{shockanglefig1.pdf}
\caption{
Left panel: The approximate bremsstrahlung continuum and heuristic
NLF distributions in arbitrary units, as a function of $\energy$ and $\ti$
respectively, showing a two-$T$ DEM model with $\temp = 1/3$ and 1 and a hotter/cooler emission-measure
ratio of 3/2, and a single-initial-$T$ CIT model
with $\tinitial = 5/4$.
The two-$T$ bremsstrahlung continuum is the thin solid curve, the single-initial-$T$ bremsstrahlung
continuum is the dashed curve that follows along it,
and the NLF distributions for lines with emissivity functions with FWHM of a factor of 3 are shown with
the thick solid and dashed curves, for the steady two-$T$ and impulsive single-initial-$T$ models respectively.
Also shown are the NLF distributions for lines with emissivity functions with FWHM of a factor of 2,
where the dot-dashed curve is for the two-$T$ DEM, and the dotted curve is for the single-initial-$T$
CIT model.  Right panel: The same curves, except showing the comparison of the difference divided by
the average for DEM and CIT models, where the thin curve is the bremsstrahlung continuum, the
thick curve is the NLF for lines with emissivity FWHM of a factor of 3 in $T$, and the dot-dashed
curve is for lines with FWHM of a factor of 2.
Again the dotted horizontal lines show the limits of 20\% uncertainty, intended as
a schematic estimate of reasonable experimental and theoretical errors.
}
\end{center}
\end{figure}
which again compares the bremsstrahlung continuum, as a function of $\energy$ (roughly keV again), and the NLF distributions as a 
function of $\ti$ for two different line emissivity widths (again FWHM of 2 and 3 in the emissivity function over $\ti$ space), all
for a two-$T$ model with $\temp = 1/3$ and 1, to a single-initial-$T$ model with $\tinitial = 5/4$.
The hot/cool emission measure ratio in the two-$T$ model is 3/2, selected to yield an almost uncanny
agreement in the bremsstrahlung continuum, despite completely different source models (one smooth and boxy, the other discrete
and bimodal).
Such close agreement between the bremsstrahlung continua in this example
suggests more generally that this continuum does not well distinguish
details in the temperature distributions of different possible source models, a conclusion that will
reappear consistently in what follows.

Since the bremsstrahlung curves in Figure 2 are equivalent to
the Laplace transform of the functions $\left ( 2\sqrt{3} \ \delta[1/x-1/3] + 3 \ \delta[1/x-1]\right )/x^{3/2}$ 
and $9.6 \ \unitstep[5/4-1/x]/(x+x^2)$, 
where $\delta[y]$ is the Dirac delta function and $\unitstep[y]$ is the Heaviside step function, this shows that the
Laplace transform of two markedly different functions can be indistinguishable even with impossibly high-precision data. 
This is possible, despite the formally unique invertibility of the Laplace transform, because of
the ill conditioning of its exponential kernel.
Thus even a small
uncertainty in the observed bremsstrahlung continuum could mix these two 
markedly different source distributions in unpredictable ways, yielding an unreliable inversion result.
The bremsstahlung continuum apparently encodes relatively few independent constraints on the source distribution,
so its ambiguity problem is especially extreme.

%This is the ill conditioning that was alluded to earlier, whereby small uncertainties in the data map into large uncertainties
%in the inferred source shape, and this is seen in Figure 2 to be especially problematic for the bremsstrahlung continuum because no
%reasonable observational precision could distinguish the two continuum spectra in that plot.

Thus it is the line fluxes in the NLF that must be relied upon to distinguish the various features of the source distribution, and the above
plots show that certain values of $\ti$ are more decisive than others for doing this, including $\ti$ below about 3/4 of the lowest 
$\temp$ included in the DEM.
The presence of emission 
from these lines is indicative of either gas cooling, or a continuous DEM, whereas its absence indicates that gas is 
continuously maintained at higher $\temp$.
Note it is not necessary that these low-$\ti$
lines actually be present in the spectrum, it is only necessary that they would be expected to be 
observed if they were indeed present, because absence of cool gas is an important signature of absence of either 
steadily cooling flows (e.g., Peterson \& Fabian 2006;
Takahashi et al. 2009), or transient cooling as well.
The above plots also show that lines at $\ti$ in the neighborhood of either the $\temp$ values in the
two-$T$ fit, or close to the centroid of those two $\temp$ values where a truly bimodal source distribution should produce a dip in the NLF,
can also help reinforce the reliability of a two-$T$ fit.

\subsection{Comparing two-$T$ DEM with two-initial-$T$ CIT fits}

The parameters of the above two-$T$ DEM model were carefully selected to yield results similar to a single-initial-$T$ model.
In general, however, thermal spectra that admit successful two-$T$ DEM fits will 
involve parameters that do not mimic any single-initial-$T$ version.
In this case, additional parameters need to be included in the CIT distribution to yield better agreement.
The closest CIT equivalent to a two-$T$ plasma is a plasma heated to two different initial $\tinitial$, and allowed to cool from
each of those.
This provides three parameters for modeling the spectral shape, just as does a two-$T$ fit, except instead of the ratio between
the emission measures, the third shape parameter is the ratio of the rates that gas is impulsively raised to the two different $T$ in the fit.
This already gives a sense of how the CIT approach brings the
parametrization closer to the actual heating physics, whenever the gas cools below its initial $\tinitial$ rather than being
continuously maintained in a DEM-type configuration.

Since the two approaches have the same number of fit parameters, they should be able to fit similarly sized sets of observed data, but
not necessarily the same ones.
For example, we expect that CIT fits will require the presence of cool gas not always present to a similar degree in a two-$T$ model, 
and we know that monotonically decreasing $F_c(\energy)$ and $F_i(\ti)$ are basic constraints on successful CIT models.
%Within these limitations, it remains to be explored when a mutual fit using both schemes is possible.
As before we expect that the bremsstrahlung continuum will prove a 
poor discriminator, and low-$\ti$ lines will be important for detecting cooling gas.
Let us further explore what formation temperatures of the heuristically treated lines can serve as effective
discriminators of these two source models.

Figure 3
%%figure 3  -- two T vs. two shock, again show extremely close bremsstrahlung and the Laplace transform ambiguity, 
%and how low-T lines or high-res lines are needed
\begin{figure}[h]
\begin{center}
\plotfiddle{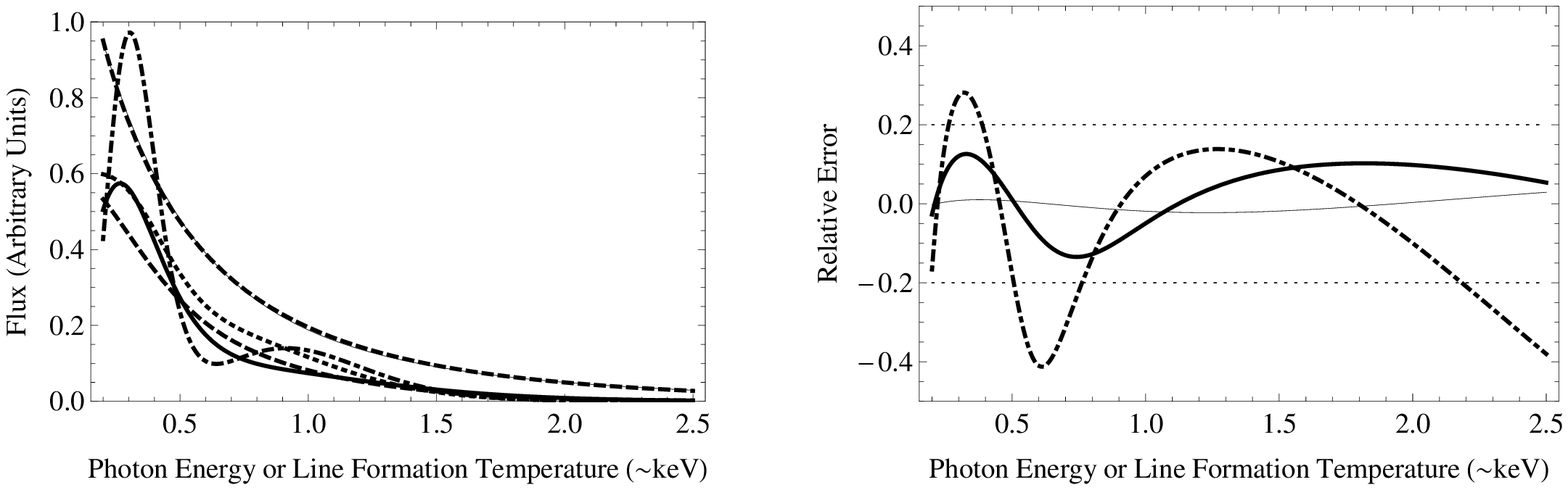}{2.6in}{-0.}{440.}{200.}{0}{0}
%\plotfiddle{PSFILE}{VSIZE}{ROTANG}{HSCALE}{VSCALE}{HTRANS}{VTRANS}
%\includegraphics[width=6.5in]{shockanglefig1.pdf}
\caption{
Left panel: The approximate bremsstrahlung continuum and heuristic
NLF distributions in arbitrary units, as a function of $\energy$ and $\ti$
respectively, comparing a two-$T$ DEM model with $\temp = 1/3$ and 1 and a hotter/cooler
emission-measure ratio of 1/2, with a two-initial-$T$ CIT model
with $\tinitial = 1/2$ and $6/5$ and a 1/2 hotter/cooler ratio of the rate that gas is impulsively heated.
The two-$T$ bremsstrahlung continuum is the thin solid curve, the two-initial-$T$ bremsstrahlung
continuum is the dashed curve that follows along it,
and the NLF distributions for lines with emissivity functions with FWHM of a factor of 3 are shown with
the thick solid and dashed curves, for the two-$T$ and two-initial-$T$ models respectively.
Also shown are  the NLF distributions for lines with emissivity functions with FWHM of a factor of 2,
where the dot-dashed curve is for the two-$T$ DEM, and the dotted curve is for the two-initial-$T$
CIT model.  Right panel: The same curves, except showing the comparison of the difference divided by
the average for DEM and CIT models, where the thin curve is the bremsstrahlung continuum, the
thick curve is the NLF for lines with emissivity FWHM of a factor of 3 in $T$, and the dot-dashed
curve is for lines with FWHM of a factor of 2.
}
\end{center}
\end{figure}
shows the example of a two-$T$ model with $\temp = 1/3$ and $1$, with half as much emission measure in the hotter gas,
and compares it with a two-initial-$T$ model with initial $\tinitial$ of 1/2 and 6/5, and half the 
relative probability of impulsively
heating to the higher $\temp$.
Once again the agreement in the bremsstrahlung continuum is very close, even though the source distributions are
drastically different in detail, while the NLF distribution provides important clues for distinguishing the source models.
Lines whose $\ti$ are below about 3/4 of the lower $\temp$ in the two-$T$ model are once again key for divining the
presence of cooling gas, and also lines with $\ti$ near the values of the two $\temp$ in the DEM model, and directly
between those two $\temp$, are also sensitive to the discrete character of the DEM.
We also see that the lines with factor 2 FWHM in their emissivity functions are much more useful for showing the
bimodality of the DEM, which is not surprising given that the factor-3 FWHM lines have difficulty diagnosing
bimodality when the $\temp$ vary by only a factor of 3.
A central lesson from this figure is that bimodality in the DEM can only be reliably diagnosed if the NLF distribution
is itself clearly bimodal, whereas a monotonicall decreasing NLF distribution is consistent with the presence of cooling
gas, but does not require it.

\subsection{Comparing two-$T$ DEM with power-law CIT fits}
% here a mutual fit is possible if not too bimodal, sufficiently monotonic in T space

Since the required monotonicity in the CIT picture always yields a smooth equivalent DEM, whenever CIT fits are ambiguous with two-$T$ fits,
it is possible that a smooth approach to parametrizing the CIT could be superior to a discrete approach to the initial $\tinitial$.
One simple smooth monotonic parametrization is a power-law form,
\beq
\label{citpower}
\CIT(\temp) \ = \ \frac{1}{1 + (\temp/\tb)^\power} \ ,
\eeq
where $\tb$ sets the characteristic scale above which a significant amount of plasma gets heated, and $\power$ sets the power law for
whatever (assumed scale-invariant) process constrains the high-temperature tail of the heating distribution.
Notice this form involves only two shape parameters for fitting the spectrum, $\tb$ and $\power$,
which is one less than a two-$T$ fit.
Thus a power-law CIT exhibits preferred simplicity
in any context where it can mimic the observable data from a two-$T$ DEM model.
When will this be possible?

Because of the presence of a significant amount of cool gas, we do not expect a power-law CIT to be able to mimic an isothermal plasma spectrum,
because a smooth CIT has even more emission, relatively speaking, at lower $\tinitial$ 
than do the single-initial-$T$ models that proved insufficient above.
However, a two-$T$ DEM that has a substantial cool-gas component may allow a power-law CIT model to mimic similar results if lines at formation
temperatures significantly lower than the lower $\temp$ in the two-$T$ model are not observationally accessible.
Figure 4
%% figure 4  
\begin{figure}[h]
\begin{center}
\plotfiddle{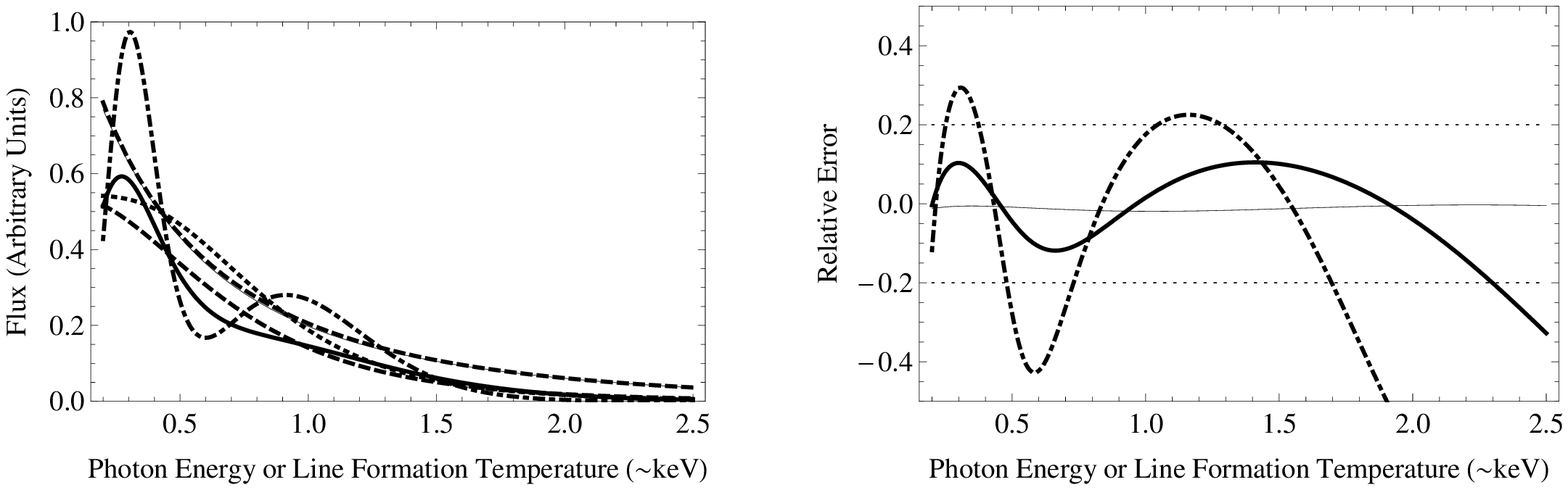}{2.6in}{-0.}{440.}{200.}{0}{0}
%\plotfiddle{PSFILE}{VSIZE}{ROTANG}{HSCALE}{VSCALE}{HTRANS}{VTRANS}
%\includegraphics[width=6.5in]{shockanglefig1.pdf}
\caption{
Left panel: The approximate bremsstrahlung continuum and heuristic
NLF distributions in arbitrary units, as a function of $\energy$ and $\ti$
respectively, comparing a two-$T$ DEM model with $\temp = 1/3$ and 1 and a hotter/cooler
emission-measure ratio of 1, with the power-law CIT model from eq. (\ref{citpower})
with $\tb = 0.95$ and $\power = 5$.
The two-$T$ bremsstrahlung continuum is the thin solid curve, the power-law CIT bremsstrahlung
continuum is the dashed curve that follows along it,
and the NLF distributions for lines with emissivity functions with FWHM of a factor of 3 are shown with
the thick solid and dashed curves, for the two-$T$ and power-law models respectively.
Also shown are the NLF distributions for lines with emissivity functions with FWHM of a factor of 2,
where the dot-dashed curve is for the two-$T$ DEM, and the dotted curve is for the power-law
CIT model.  Right panel: The same curves, except showing the comparison of the difference divided by
the average for DEM and CIT models, where the thin curve is the bremsstrahlung continuum, the
thick curve is the NLF for lines with emissivity FWHM of a factor of 3 in $T$, and the dot-dashed
curve is for lines with FWHM of a factor of 2.
}
\end{center}
\end{figure}
shows an example of a two-$T$ DEM with $\temp = 1/3$ and 1, with equal emission measure in both components, and a power-law
CIT of the form in eq. (\ref{citpower}), with $\tb = 0.95$ and $\power = 5$.
As usual, the overall scale is arbitrary, it is the shapes of the distributions we are comparing, and once again the
bremsstrahlung continuum is completely unable to distinguish these two source models.
Other familiar attributes emerge, including the importance of lines with $\ti$ below about 3/4 of the lower $\temp$ in the DEM model,
and lines with $\ti$ near either of the two $\temp$ in the DEM, or the midpoint between them.
Once again, bimodality is only clearly visible in the NLF distribution for lines with narrow emissivity functions, and if such
diagnostically powerful lines do not happen to have $\ti$ values in the critical regimes, considerable ambiguity between the two
highly different source models is possible.
If it does happen that lines with appropriate $\ti$ are not accessible in the dataset, we also note that a fit with only two shape
parameters, the power-law CIT model, can sometimes fit data that could also be modeled with the three shape parameters of a two-$T$ DEM.

\subsection{Discussion}
%ambiguity, CIT can fit with 2 parameters instead of 3, edges of spectrum are important but hard to detect

Of course, all that has been shown above are several illustrative examples, yet a consistent picture emerges in regard to 
distinguishing situations where one- and two-$T$ DEM fits succeed, and those where CIT-type models could succeed as well.
The overall conclusions are that the bremsstrahlung continuum cannot distinguish discrete DEM models from CIT models, but lines
with $\ti$ below about 3/4 of the lowest $\temp$ in the DEM model can do so easily, if such lines are accessible.
Also, lines with emissivity functions with FWHM of about a factor of 3 in $\ti$ space are not sufficient to diagnose features,
such as bimodality, in
the DEM over a similar $\temp$ range, and since many lines are found to have FWHM in the range 2--3 in their
emissivity functions, bimodality in the DEM over a similar $\temp$ range requires careful selection of the lines used,
a point investigated in more detail in Gayley et al. (2014).
Also, to be effective in constraining features on the DEM on scales like a factor of 3 in $\temp$, the lines must
not only have narrow emissivity, they must also have $\ti$ values near the desired features in the DEM $\temp$.

It bears noting that because we have seen that 
lines with $\ti$ above about 3/4 of the lowest $\temp$ in the DEM fit cannot diagnose the presence or
absence of plasma cooling below that $\temp$,
any time a discrete DEM fit is arrived at, and the lowest $\temp$ in the DEM is at least 4/3 times the lowest $\ti$ of the
lines that the spectrum has access to, the resulting discrete character of the DEM is not robustly inferable.
In light of the fact that X-ray spectrometers often lose sensitivity at long wavelengths, and foreground absorption can
extinct long-wavelength emissions as well, a tendency for low-$\ti$ lines to appear at longer wavelengths presents a problem.
It is clear that lines with low $\ti$, yet short wavelengths, are especially valuable for diagnosing cooling gas, 
especially if they combine
those attributes with an emissivity width of roughly a factor of 2 in $\temp$.
Iron L-shell lines are numerous in thermal X-ray spectra, and often have relatively narrow emissivities, so the shortest wavelength
of these lines seem particularly important in this context.

Lines and continua with high $\ti$ or high $\energy$ can also be used as discriminators, but are not included here because the
signal is usually weak.
Thus, in general the diagnostics of greatest
discriminiting value often appear near the edges of the distribution, where they are most
difficult to obtain.
This suggests that spectrographs that are sensitive over a wide dynamical range and 
with wide energy coverage may be of particular importance.
Note that it is not necessarily the lines with the highest flux that matter most, both because such lines may be strong
expressly because they have broad emissivity functions, and also because it may be an upper limit on a null detection
that is more important for distinguishing DEM and CIT type models.

With all the above issues in mind, let us now consider several thermal X-ray sources selected from across astronomy, and see what
insights this heuristic analytic approach can yield when applied to datasets that have been fit with two-$T$ DEM models.
Of particular importance is the issue of when a two-$T$ DEM fit can be used as evidence for a truly bimodal source mechanism, and
when is it simply a reflection of the chosen modeling domain.

\section{Basic Applications to Two-$T$ DEM Fits Across Stellar Astronomy}
%% The goal here is to draw some basic conclusions about when bimodal DEMs have been inferred, and under what circumstances those
%%   could also be modeled by a CIT.  A key issue is whether or not there are low_Ti lines constrained by the spectrum.

Now that a general scheme has been developed to estimate the shape of the bremsstrahlung continuum and the NLF distribution from
the shape of a DEM or CIT distribution, we can explore the insights it brings to real astrophysical applications.
For example, we can ask if any given DEM that was found to produce a satisfactory fit to a spectrum could alternatively admit
a CIT-type fit.
As a general proof of concept, here the focus will be strictly on two-$T$ DEM fits, which were prevalent prior to the most recent
generation of X-ray instrumentation, and retain their relevance in regard to multitemperature fits, taken two temperatures
at a time.
It seems that as the instruments have become more precise and signal-to-noise has improved, the 
standard fit type has progressed from isothermal, to
two-$T$, to multithermal, and ultimately to continuous DEM models, and the results found here may give some indication as to why that is.
We must remember that the entire justification for modeling the DEM, even though by itself it is not a quantity that is well adapted
for understanding heating mechanisms, is that it is viewed as the model-independent generator of the X-ray spectrum for plasmas in CIE.
But if bimodality is merely just one from a menu of equally successful fit options,
then a two-$T$ DEM cannot be regarded as model independent, and a primary goal of using a DEM parametrization is compromised.
Let us critically examine whether or not this issue could be creating problems in real applications, using the heuristic tools
derived above.

\subsection{X-rays from galaxy clusters}
%   --Centaurus cluster (Takahashi, I., Kawaharada, M., Makishima, K., Matsushita, K., Fukazawa, Y., Ikebe, Y., Kitaguchi, T.,
%Kokubun, M., Nakazawa, K., Okuyama, S., Ota, N., \& Tamura, T. 2009, ApJ, 701, 377.
%  they say " The 1.1 keV component has turned out to be weak in contrast to the two
% adjacent ones, implying that the significantly detected two components are discrete entities rather
% than representing a continuous temperature distribution. "  But note the T ratio is only a factor of 2, how can they say this?
%  Their T ratio is 2.25, like 0.8 and 1.7 keV.
%   Note the 1 sigma errors range from 3% to 8%, but that's just the observational error, there's also theory uncertainties, so 10-20% may be reasonable
%  Still, they make a good case that there is a paucity of low-T line emission, suggesting it is neither a cooling flow nor shocks that cool radiatively,
%  but rather the gas is maintained consistently above 0.5 keV, perhaps by cool loops.
%  I would point out that perhaps the low-T gas mixes with even cooler gas, so gas may be lost from the cooling stream as time goes on.
%  Still, it suggests the CIT approach won't work to T < 0.7 keV or so.
%   I think a key point to stress is that you can only tell that cooling flow, or CIT parametrizations in general, fail if you have low-T lines that
%  you are sure are not seen in the spectrum and should be, i.e., their absence cannot be explained by foreground absorption or low-T mixing

Galaxy clusters are sources of diffuse X-ray emission, a fact that was originally interpreted as being due to cooling flows of
shocked and virialized inflowing gas (e.g., White \& Rees 1978; White \& Frenk 1991) 
%% White, S. D. M. \& Rees, M. J. 1978, MNRAS, 183, 341
%% White, S. D. M. \& Frenk, C. S. 1991, ApJ, 379, 52
with possible feedback modifications (e.g., Croton et al. 2006).
%% Croton, D. J., Springel, V., White, S. D. M., De Lucia, G., Frenk, C. S., Gao, L., Jenkins, A., Kauffmanm, G., Navarro, J. F., \& Yoshida, M. 2006, MNRAS, 365, 11
This interpretation requires modifications for understanding the absence of large amounts of cooling gas (Peterson et al. 2003), 
%% Peterson, J. R., Kahn, S. M., Paerels, F. B. S., Kaastra, J. S., Tamura, T., Bleeker, J. A. M., Ferrigno, C., \& Jernigan, J. G. 2003, ApJ, 590, 207
an absence that also presents a challenge
to CIT type models, which include cooling gas similar to what a cooling flow would produce, albeit stochastic rather than steady.
Restricting our analysis here to cases where two-$T$ DEM models have been recently applied, we note that Takahashi et al. (2009)
find that the central (cD) region of the Centaurus cluster yields thermal X-rays that submit to a two-$T$ model with $\temp \cong 0.77$
and $\temp \cong 1.9$, in units where $\To = 1$ keV is a reasonable approximation of the crossing between line and bremsstrahlung cooling.
Their model allowed an intermediate $\temp \cong 1.1$, but they found they did not need this component to fit the spectra, and they took
this as evidence that the source distribution is truly bimodal.
The bimodality in turn led to a picture whereby cooler gas was interspersed with hotter gas in a two-phase, fixed-pressure configuration.
Hence, their analysis involved parametrizing a multithermal DEM in a manner that is often advertised as ``model independent,'' as that is
the only reason to model a DEM in the first place, and the resulting fit was taken as evidence for a particular physical model.

The difficulty with this approach is brought out by the above heuristic analysis, which suggests that the fundamental line diagnostic,
the NLF distribution, cannot take lines with emissivity FWHM of factors of 2--3 in $T$, and use them to deduce 
unambiguous structure in the DEM on
$T$ scales less than that FWHM (such as the differences between $\temp = 0.78$ and 1.1, or between 1.1 and 1.9).
So although the model inferred by Takahashi et al. (2009) is a valid way to interpret the data, it is not a model independent conclusion
that the X-ray spectrum must be interpreted in terms of gas that is being maintained continuously in one of two different $T$ domains.
Tests of whether or not the X-ray spectrum requires, rather than allows, such a bi-phase interpretation would be more difficult than a
successful spectral fit,
and might even be impossible if that information is simply not present in an X-ray spectrum, if the ill conditioning limits the
availability of the necessary independent constraints.
Hence, alternative heating models, such as the impulsively heated CIT picture, or the related cooling-flow model,
are not necessarily informed by the success of a two-$T$ model, and fits of the latter type could be separately attempted
(perhaps informed by some of the issues raised in Peterson \& Fabian 2006).

This point is brought out more clearly in Figure 5,
% figure 5
\begin{figure}[h]
\begin{center}
\plotfiddle{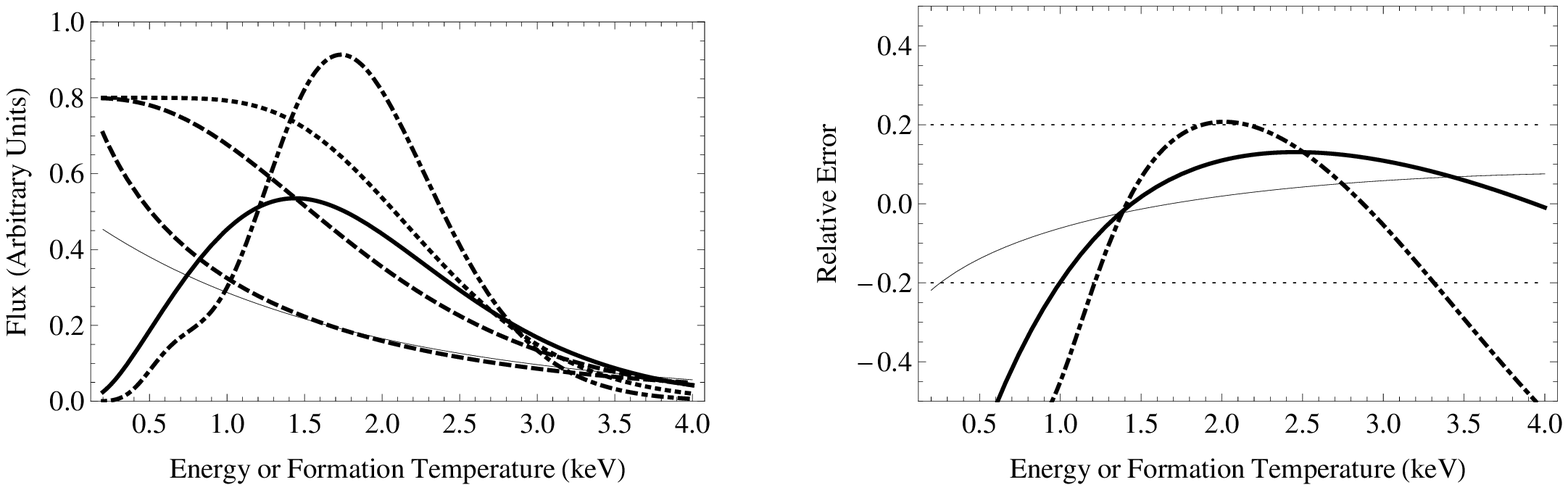}{2.6in}{-0.}{440.}{200.}{0}{0}
%\plotfiddle{PSFILE}{VSIZE}{ROTANG}{HSCALE}{VSCALE}{HTRANS}{VTRANS}
%\includegraphics[width=6.5in]{shockanglefig1.pdf}
\caption{
Left panel: The approximate bremsstrahlung continuum and heuristic
NLF distributions in arbitrary units, as a function of $\energy$ and $\ti$
respectively, comparing a two-$T$ DEM model with $\temp = 0.77$ and 1.9 and a hotter/cooler
emission-measure ratio of 16, with a single-initial-$T$ CIT model
with $\tinitial = 2.6$.
The curve conventions are as in the above figures, as is the right panel.
}
\end{center}
\end{figure}
where the heuristic bremsstrahlung continuum and NLF distributions are plotted using the expressions
in this paper, for a two-$T$ model
with $\temp = 0.77$ and 1.9, and a hot/cool emission-measure ratio of 16.
No signature of bimodality survives in either the bremsstrahlung continuum or the NLF distribution, even for
lines with narrow emissivities of a FWHM of a factor of 2 in $\ti$, so a two-$T$ DEM fit could always be obtained
with a smooth DEM as well. 
The absence of bimodality in the NLF here is due in part to the weakness of the cooler component, yet
even when the cool emission measure is not small,
as in Figure 6 which roughly corresponds to the Takahashi et al. (2009) model summed over shells outside
the central (cD) region of the Centaurus cluster, a clear signature of bimodality is still not seen, because
the ratio of the two $T$ is not large.
In the model shown in Figure 6, 
%% figure 6
\begin{figure}[h]
\begin{center}
\plotfiddle{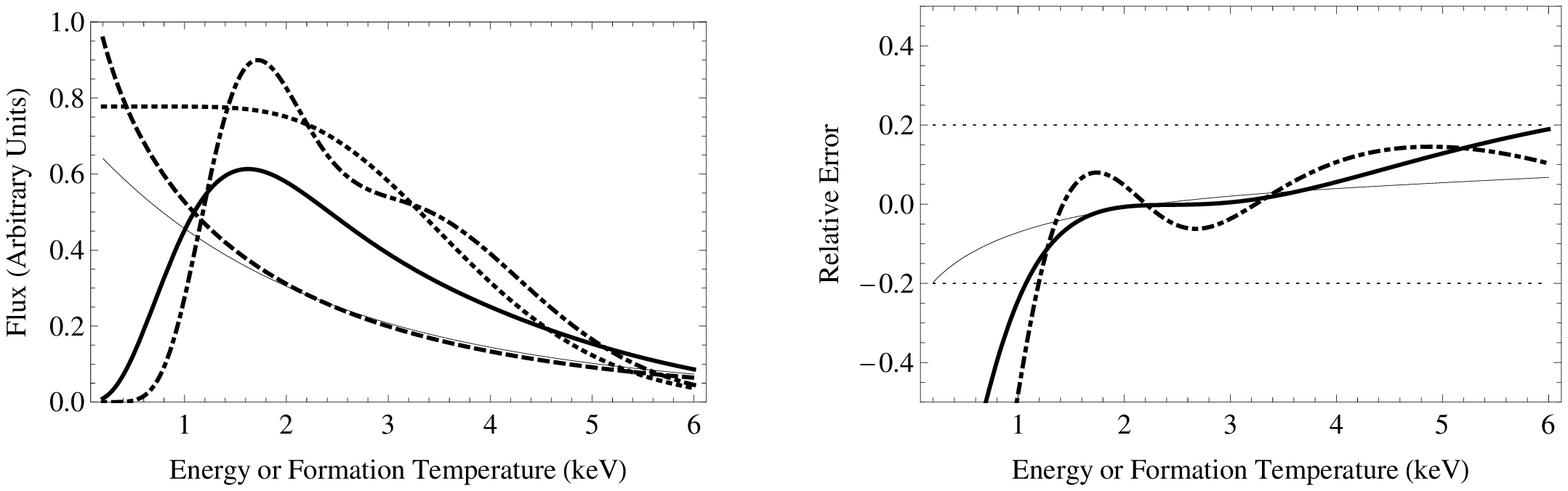}{2.6in}{-0.}{440.}{200.}{0}{0}
%\plotfiddle{PSFILE}{VSIZE}{ROTANG}{HSCALE}{VSCALE}{HTRANS}{VTRANS}
%\includegraphics[width=6.5in]{shockanglefig1.pdf}
\caption{
Left panel: The approximate bremsstrahlung continuum and heuristic
NLF distributions in arbitrary units, as a function of $\energy$ and $\ti$
respectively, comparing a two-$T$ DEM model with $\temp = 1.8$ and 3.8 and a hotter/cooler
emission-measure ratio of unity, with a single-initial-$T$ CIT model
with $\tinitial = 4.2$.
The curve conventions are as above, as is the right panel.
}
\end{center}
\end{figure}
$\temp = 1.8$ and 3.8 with a hot/cool emission-measure ratio
of unity, and since the 
%The Takahashi et al. (2009) results found the emission measure of the hotter and cooler components varied with radius,
%but on the whole were similar, so Figure 6 assumes a unity ratio of the hot/cool emission measures.
discreteness in the DEM model is still not apparent in the observables, spectral information by itself cannot
unambigusouly establish discreteness in the temperature structure anywhere in the Centaurus cluster.

Figure 6 also shows
that when the hot and cool components have similar emission measure, it is difficult to distinguish plasma
that is always maintained above some minimum $T$ (in this case $\temp = 0.77$), from a CIT or cooling-flow type distribution
that produces even cooler gas, unless lines with $\ti$ below about 3/4 of the minimum $\temp$ are accessible in the spectrum
(so fall within the wavelength window of the instrument, and are not quenched by foreground absorption or suppressed by low 
abundance).
Certainly the bremsstrahlung continuum is seen to be of decidedly limited usefulness in drawing such distinctions, unless it can be
reliably constrained at low $\energy$.

The paucity of low-$T$ gas, relative to a cooling-flow model, is even more pronounced in the central (cD) region of the
Centaurus cluster, as seen above in Figure 5.
Here the Takahashi et al. (2009) fit is similar to a two-$T$ fit by Sanders et al. (2008), which is compared
in Figure 5 with a simpler CIT model with only one shape parameter, the single initial $\tinitial = 2.6$.
It may be seen from the figure that once again,
only lines with narrow emissivity functions at the crucial $\ti$ values, or lines with low $\ti$ values,
can distinguish a two-$T$ DEM distribution from a CIT distribution. 
When the hotter component is so dominant,
the $\ti$ regime where differences appear shift to roughly 3/4 of the {\it higher} $\temp$, rather than 3/4 of the {\it lower}
$\temp$ as when the cooler component is more significant.
%%  ref Sanders, J. S., Fabian, A. C., Allen, S. W., Morris, R. G., Graham, J., \& Johnstone, R. M. 2008, MNRAS, 385, 1186.

Apparenly we may conclude 
that it is crucial to assess whether or not reliable line fluxes, or reliable line-flux upper limits, can be obtained at 
low $\ti$ to determine if it is indeed true that insufficient cool gas is present to allow
a CIT-type model, or a cooling-flow model, to succeed.
It appears that fits to the central regions of the Centaurus cluster are more clearly hostile
to a cooling-flow
interpretation
(or more generally, an impulsively heated CIT distribution) than are the outer shells of the cluster.
%Foreground absorption is a crucial difficulty in this regard (ref), since it tends to obscure long-wavelength lines that
%might otherwise provide excellent low-$T$ constraints.
But even if the presence of cooling gas is indeed ruled out, the cooling 
gas might be 
shunted into some nonradiative cooling channel, such as the mixing with previously cooled
gas that may occur in radiative shocks (Owocki et al. 2013), 
%% Owocki, S. P.; Sundqvist, J. O.; Cohen, D. H.; Gayley, K. G. 2013, MNRAS, 429, 3379
necessitating modifications to the $f_o(\temp)$ term in 
CIT models to allow for weaker soft X-ray emission.
Or, if the absence of cooler emission is taken as evidence that the gas is being continuously maintained at higher $T$,
it still does not necessarily require a bimodal interpretation of the source
distribution in the Centaurus cluster, since
Figures 5 and 6 show that the NLF distributions for lines of various emissivity widths have difficulty establishing unambiguous
bimodality for $T$ ratios barely above 2.
Alternative approaches to simple multithermal DEM modeling may be required to probe these questions more deeply.

\subsection{Diffuse galactic X-rays}
%%  usher in the prospects for spatial separation of two-T components, hotter is beyond GMC, though note that isn't necessarily
%%   two discrete T, just two combined DEMS with different <T>, so could be bi-initial-T as much as bi-continuous-T
%% Kuntz

Closer to home, in our own galaxy, the diffuse X-ray emission is sometimes modeled with a two-$T$ DEM.
Kuntz \& Snowden (2008) find 0.2 and 0.6 keV components toward an absorbing cloud about 3 kpc away along $l=111 \deg,$
with a hot/cold emission-measure ratio of 9.
Figure 7
%% figure 7
\begin{figure}[h]
\begin{center}
\plotfiddle{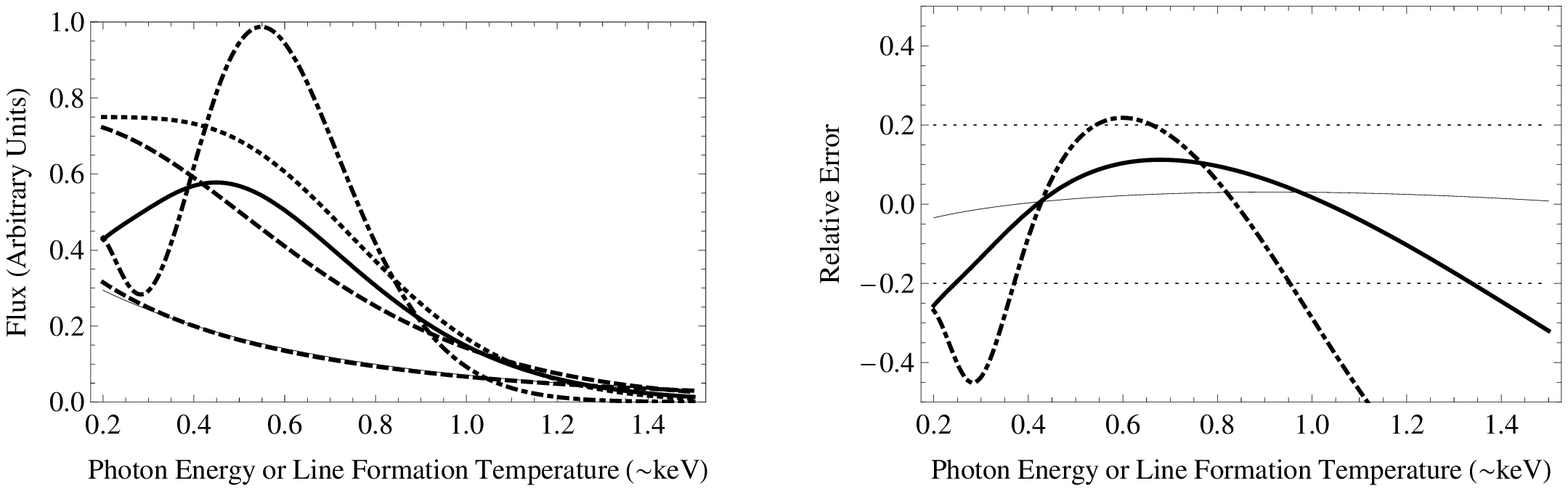}{2.6in}{-0.}{440.}{200.}{0}{0}
%\plotfiddle{PSFILE}{VSIZE}{ROTANG}{HSCALE}{VSCALE}{HTRANS}{VTRANS}
%\includegraphics[width=6.5in]{shockanglefig1.pdf}
\caption{
Left panel: The approximate bremsstrahlung continuum and heuristic
NLF distributions in arbitrary units, as a function of $\energy$ and $\ti$
respectively, comparing a two-$T$ DEM model with $\temp = 0.2$  and 0.6 and a hotter/cooler
emission-measure ratio of 8.9, with a single-initial-$T$ CIT model
with $\tinitial = 0.9$.
The curve conventions are as above, as is the right panel.
}
\end{center}
\end{figure}
shows the heuristic NLF distribution and bremsstrahlung continuum, in comparison to a single-initial-$T$ CIT model
with $\temp = 0.9$.
Again the continuum is of no use in distinguishing the models, but lines at low $\ti$ are capable of doing so.
To determine bimodality, lines with narrow emissivity functions are needed with $\ti$ in the vicinity of the peaks at
0.2 and 0.6 keV, and the valley in between as well.
Only if such reliable lines are accessible, and have well-constrained abundances, can structure in the DEM be unambiguously 
inferred from spectral information alone.
However, Kuntz \& Snowden (2008) find that the 0.2 and 0.6 keV components are separated by an
absorbing cloud, which provides supplemental spatial information for establishing the bimodality.
The potential for ambiguity in the absence of spatial information is made clear by models of M101, which
originally found that a two-$T$ fit was successful (Kuntz \& Snowden 2003),
but later found that the fit did not survive better data (Kuntz \& Snowden 2010).
This type of ambiguity raises the potential of considering an altogether different parametrizations, such as a CIT type,
better tailored to whatever heating hypothesis is being investigated.
In other words, the hope of achieving model-independent results may be unwarranted.

\subsection{RSCVn systems}
%   --RSCVn (Gehrels and Williams, Swank et al 1981)
%   note Sanz-Forcada, Brickhouse, \& Dupree 2003  point out they neglect conduction, I'd add they neglect force balance altogether
%     also, Sanz-FOrcada et al find different DEM peaks, though they use a continuous distribution based on Jordan technique and
%                           settle for flux predictions that are off by sometimes a factor of 2 and worse.
%   also, see Huenemoerder, Canizares, \& Schulz  2001 for the checkered history of seeing discrete
%         low-T components in II Pegasi, which might be quite reminiscent of Capella from Swank to Gu

Turning our attention to stars, it has long been known that rapid rotation correlates with X-ray emission (Walter \& Bowyer).
%% Walter, F. M., \& Bowyer, S. 1981, ApJ, 245, 671
Early efforts to model the coronae of RS CVn stars, which are spun up by binary tidal effects, 
often involved two-$T$ fits (e.g., Swank et al. 1981),
%%ref Swank, J. H., Holt, S. S., White, N. E., \& Becker, R. H. 1981, ApJ, 246, 208
a fact that was even given a physical interpretation
as being caused by two domains of thermal stability (Gehrels \& Williams 1993).
%%ref  Gehrels, N. \& Williams, E. D. 1993, ApJL, 418, L25
However, there are many physical reasons why we should not expect thermal stability domains to induce a bimodal DEM involving
gas that is continuously maintained in those stable domains.
First of all, stability is a necessary condition for maintaining gas at a fixed $T$, but it is not sufficient, because there
also must be an energy balance in the first place.
To allow the gas to remain in a narrow temperature window, the heating would also have to be regulated to balance the radiative
cooling function in the narrow domains of stability.
In other words, temperature stability does not just require the $T$ to lie in a regime of positive slope in the radiative
cooling function, it also requires that the ratio of heating per
particle to density must maintain the necessary magnitude to fall into that same regime, and the latter has no
reason to be regulated the way the former could.
Furthermore, as pointed out by Sanz-Forcada et al. (2003), 
%% ref   Sanz-Forcada, J., Brickhouse, N. S., \& Dupree, A. K. 2003, ApJS, 145, 147
such a stability argument ignores thermal conduction, so would
not be applicable to maintaining gas in the solar corona within particular stable $T$ regimes.

Now, despite these theoretical objections, if the spectra unambiguously require bimodal DEM, then it can be taken as true,
and theorists must solve the puzzle somehow.
For example, the fact that RS CVn stars are binaries might also lead us to wonder if bimodal DEMs might be associated
with emission from each of the two stars.
But if ill conditioning in the fitting procedure implies that a bimodal fit is just one of several possible approaches,
then the tendency for the models to avoid the regions of thermal instability, or the tendency to associate spatially
distinct emission regions with distinct temperatures, might just be a coincidence of the
general similarities in a modeling approach that is expressly looking for two-$T$ fits.
If so, the above theoretical objections seem to caution against such a modeling choice, because most heating mechanisms
would seem more likely to produce a range of $T$.
Still, it is not so much the $T$ range that is the crucial issue, as any DEM could be viewed as primarily schematic,
a more fundamental science question would seem
to be whether or not the plasma in RS CVn coronae is maintained at locally constant $T$,
or if the heating and cooling are locally impulsive and transient.
Fitting efforts using a CFT approach should be able to address the latter possibility.

As an example of the method, let us consider recent spectral modeling efforts for the RS CVn star II Peg.
Mewe et al. (1997) 
%% ref   Mewe, R., Kaastra, J. S., ven den Oord, G. H. J., Vink, J., \& Tawara, Y. 1997, A\&A, 320, 147
find a bimodal fit, with two temperatures near 0.9 and 1.9 keV, and an emission measure ratio of
0.7, hotter to cooler.
The above analysis suggests suspicion of bimodality on any temperature scale that is as fine as the
typical FWHM of the emissivity functions of the lines used to create the fit, so we might already be
inclined to suspect that the bimodality in this example is at best ambiguous.
Figure 8 
%% figure 8  
\begin{figure}[h]
\begin{center}
\plotfiddle{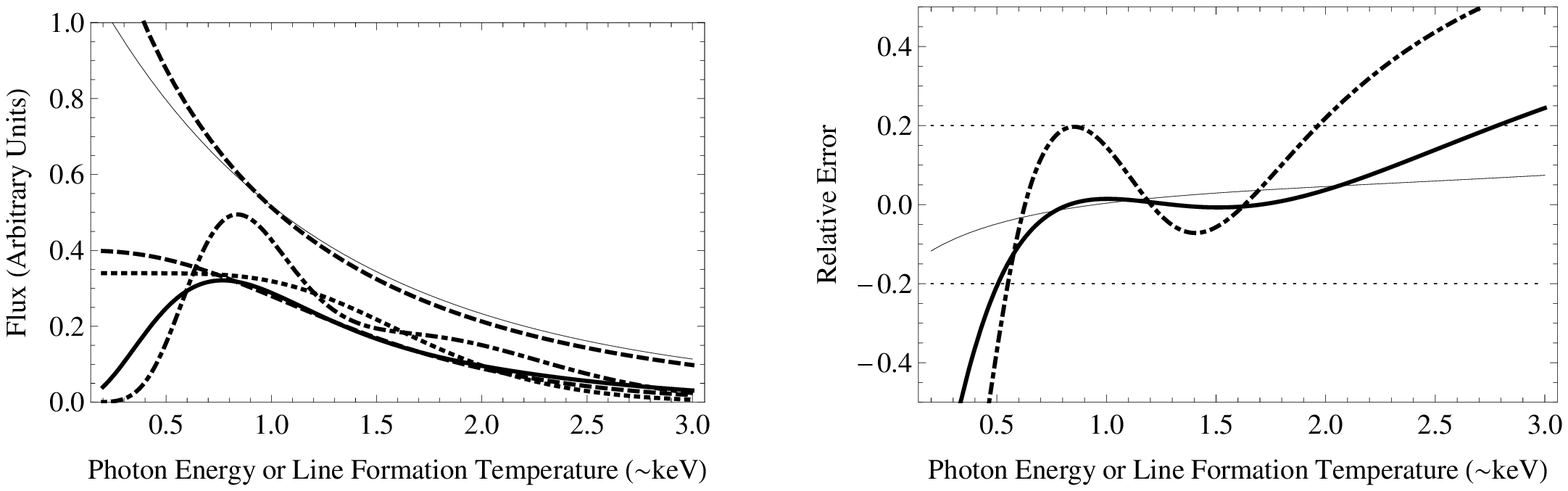}{2.6in}{-0.}{440.}{200.}{0}{0}
%\plotfiddle{PSFILE}{VSIZE}{ROTANG}{HSCALE}{VSCALE}{HTRANS}{VTRANS}
%\includegraphics[width=6.5in]{shockanglefig1.pdf}
\caption{
Left panel: The approximate bremsstrahlung continuum and heuristic
NLF distributions in arbitrary units, as a function of $\energy$ and $\ti$
respectively, comparing a two-$T$ DEM model with $\temp = 0.9$ and 1.9 and a hotter/cooler
emission-measure ratio of 0.7, with a single-initial-$T$ CIT model
with $\tinitial = 1.9$.
The curve conventions are as above, as is the right panel.
}
\end{center}
\end{figure}
reinforces this suspicion, by showing that the NLF distribution that could be fit by such a two-$T$ 
source model does not exhibit any clear bimodality, so does not require a bimodal source, even for lines with
emissivity FWHM of a factor of 2 in $\ti$.
The figure also shows that only the lowest-$\ti$ lines, again those below about 3/4 of the lowest $\temp$ in the
model (which  here is roughly $\temp \cong 0.9$, given the approximate connection between the $\temp$ scale
and the crossover between line and bremsstrahlung cooling at around 1 keV), can clearly distinguish the 
three-shape-parameter two-$T$ model
from a one-shape-parameter single-initial-$T$ CIT with initial $\temp = 1.9$.
Thus we have a one-parameter model giving some similar results to a three-parameter model, where the latter includes
detailed structure that the observations are simply not suitably well conditioned to be able to confirm.

Even more importantly, a more recent fit by Huenemorder et al. (2001)
%% ref  Huenemorder 2001
to HETG data give the results shown in Figure 9,
%% figure 9    Huenemorder 2001 DEM for II Peg
\begin{figure}[h]
\begin{center}
\plotfiddle{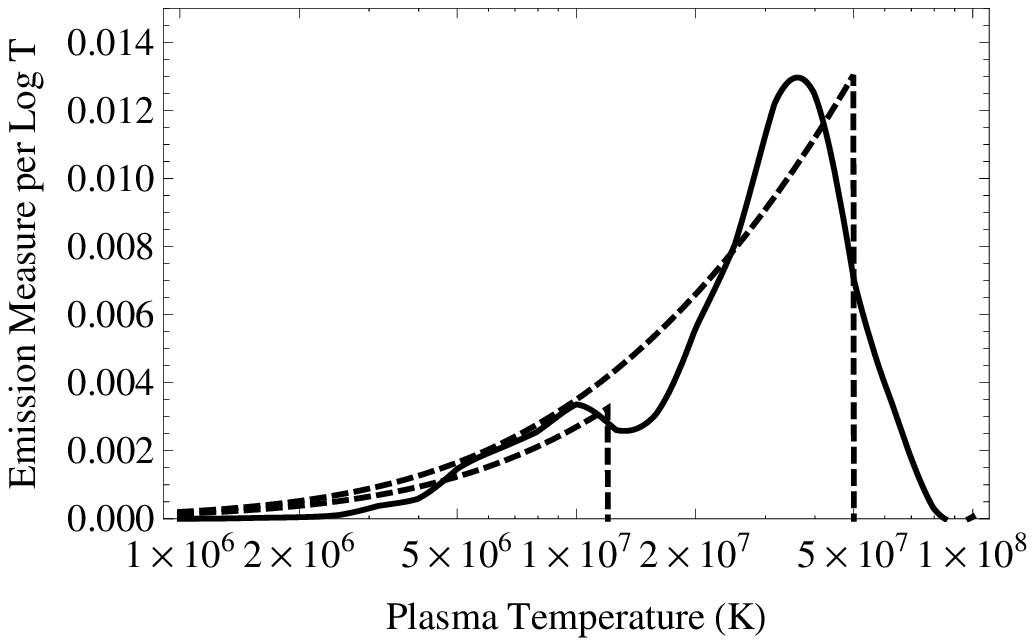}{2.6in}{-0.}{300.}{200.}{0}{0}
%\plotfiddle{PSFILE}{VSIZE}{ROTANG}{HSCALE}{VSCALE}{HTRANS}{VTRANS}
%\includegraphics[width=6.5in]{shockanglefig1.pdf}
\caption{
The solid curve is the DEM (per unit log $T$ rather than per unit $T$)
for II Peg inferred by Huenemoerder et al. (2001), and the dashed curves are the
DEM that are equivalent to a single-initial-$T$ CIT with initial $T = 12$ and 50 MK,
for pure radiative cooling using the heuristic approximations above.
The approximate connections between the two CIT models and the two DEM components suggests the CIT model that would fit the
data would have gas be introduced to the impulsive heating at about the same rate in the two components
but to different initial $T$.
Whether or not there are really two components there would require more detailed analysis, but temporal (flare) variations
suggests this is indeed the case.
}
\end{center}
\end{figure}
which involves a smooth DEM with only minor bimodality, and the bimodality that exists appears to be due to whether or not
a flaring component is manifested in the spectrum at the time of observation.
Figure 9 also shows the equivalent DEM (per $\log T$) from eq. (\ref{conversion}) 
that would be obtained from a single-initial-$T$ CIT model with
initial $\temp$ of either 1 or 4.3 (about 12 MK or 50 MK if the line/bremsstrahlung crossover is near 1 keV),
corresponding to the quiescent and flaring components respectively.
This result suggests that the X-ray spectrum from II Peg might be able to be fit by a CIT model whereby in quiescent
periods, gas is impulsively heated to a range of $T$ around 12 MK and allowed to cool transiently through all lower $T$, 
and in flaring periods, the rate that gas is introduced to the impulsive heating is essentially
the same, but it is raised to a higher initial $T$ around 50 MK.
Also note that a similar type of fit was found successful for a class of cataclysmic variables that is not
highly photoionized, interpreted by Mukai et al. (2003) in terms of a steady accretion flow, which is difficult
to distinguish from an impulsive
CIT-type distribution with a single initial $T$.
%%ref   Mukai, K., Kinkhabwala, A., Peterson, J. R., Kahn, S. M., \& Paerels, F. 2003, ApJL, 586, L77.

The Swank et al. (1981) bimodal DEM fits, interpreted by Gehrels \& Williams (1993) in terms of two regions of thermal
stability, often involved $T$ ratios of around 5, not the factor 2 seen in the Mewe et al. (1997) model for II Peg,
so lines with appropriate $\ti$ and correct abundances should be able to determine in the DEM is indeed bimodal over such
a wide $T$ contrast.
This is supported by Figure 10,
%% figure 10
\begin{figure}[h]
\begin{center}
\plotfiddle{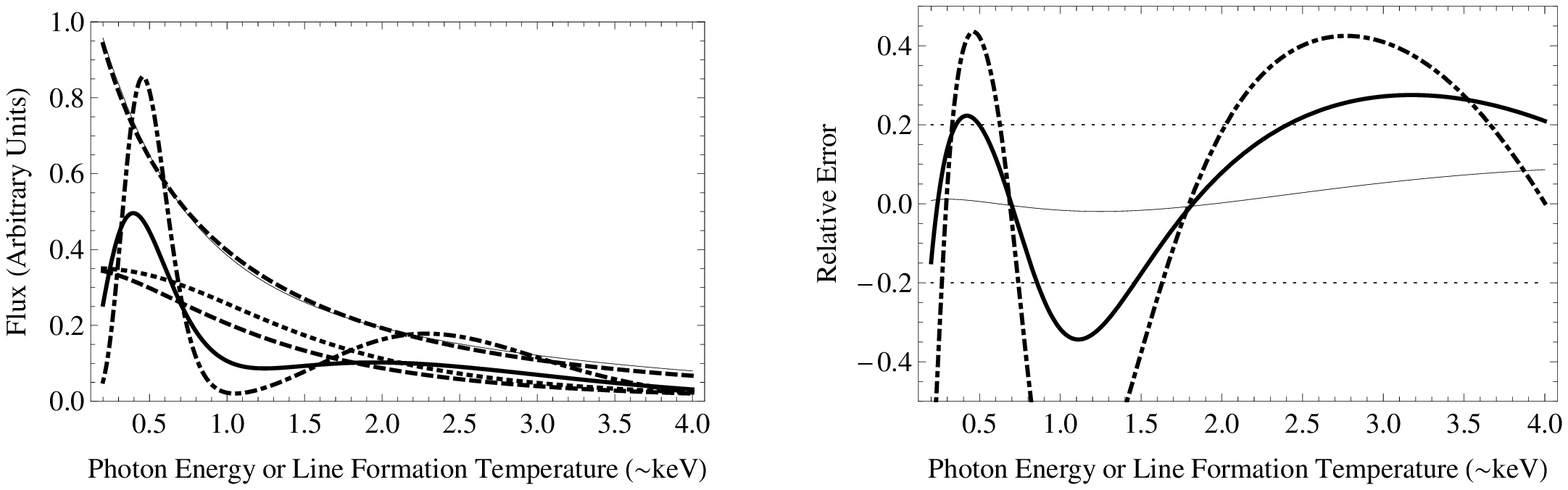}{2.6in}{-0.}{440.}{200.}{0}{0}
%\plotfiddle{PSFILE}{VSIZE}{ROTANG}{HSCALE}{VSCALE}{HTRANS}{VTRANS}
%\includegraphics[width=6.5in]{shockanglefig1.pdf}
\caption{
Left panel: The approximate bremsstrahlung continuum and heuristic
NLF distributions in arbitrary units, as a function of $\energy$ and $\ti$
respectively, comparing a two-$T$ DEM model with $\temp = 0.5$ and 5 and a hotter/cooler
emission-measure ratio of unity, with a power-law CIT model from eq. (\ref{citpower}) with
$\tb = 1.7$ and $\power = 3$.
The curve conventions are as in Figs. 1--8, as is the right panel.
}
\end{center}
\end{figure}
which plots as usual the approximate bremsstrahlung continuum and NLF distributions for a two-$T$ model, here with $\temp = 0.5$
and 2.5, and equal emission measures in the two components.
However, the figure also shows that
if lines are not present at the necessary $\ti$, or if the abundances are incorrectly skewed to support the 
bimodality, smoother CPT-type fits that involve impulsive heating to a range of $T$ around 1.7 keV might also be possible.

The possible presence or absence of bimodal character in RS CVn DEM fits is further brought out by
additional RS CVn smooth DEM fits in Huenemoerder et al. (2013) for HR 1099 and 
$\sigma$ Gem.
Applying the heuristic analysis here to those DEM results yields Figures 11 and 12, 
%% figure 11 figure 12
\begin{figure}[h]
\begin{center}
\plotfiddle{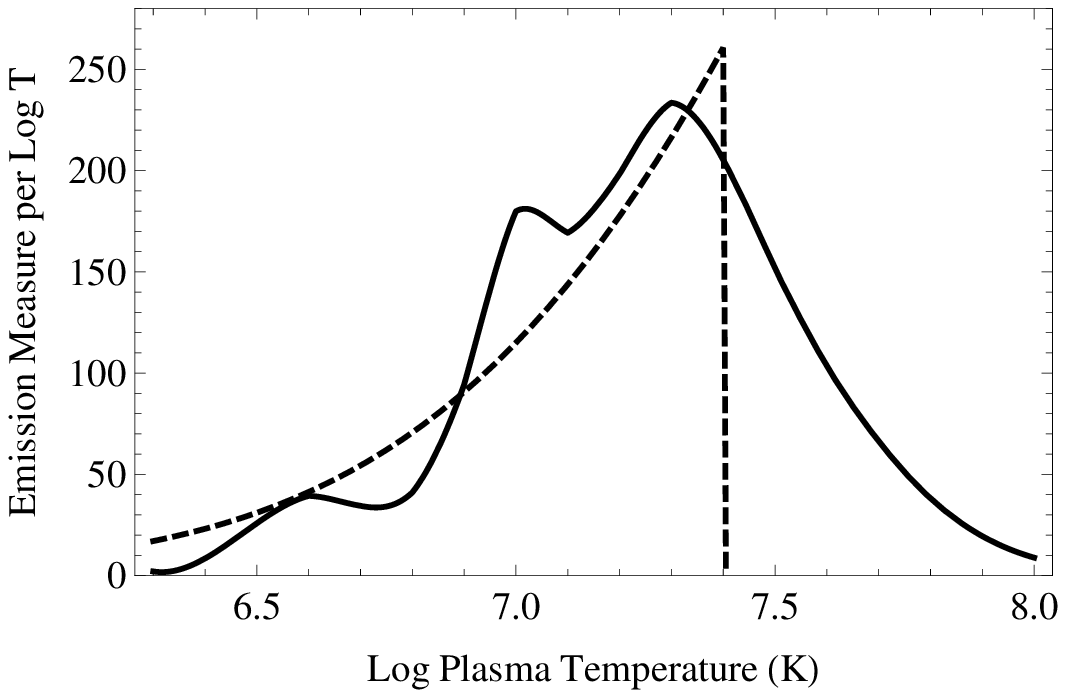}{2.6in}{-0.}{300.}{200.}{0}{0}
%\plotfiddle{PSFILE}{VSIZE}{ROTANG}{HSCALE}{VSCALE}{HTRANS}{VTRANS}
%\includegraphics[width=6.5in]{shockanglefig1.pdf}
\caption{
The solid curve is the DEM (per unit log $T$ rather than per unit $T$ as elsewhere in this paper)
for HR 1099 inferred by Huenemoerder et al. (2013), and the dashed curve is the
DEM that is equivalent to a single-initial-$T$ CIT with initial $T = 25$ MK,
for pure radiative cooling using the heuristic approximations above.
}
\end{center}
\end{figure}
where it is seen that HR 1099 may also
admit to a general interpretation in terms of impulsive heating to a range in $T$ above about 20 MK, whereas $\sigma$ Gem
may require a bimodal distribution, either of continuously maintained $T$ mapped out by the DEM, or of two separate impulsively
\begin{figure}[h]
\begin{center}
\plotfiddle{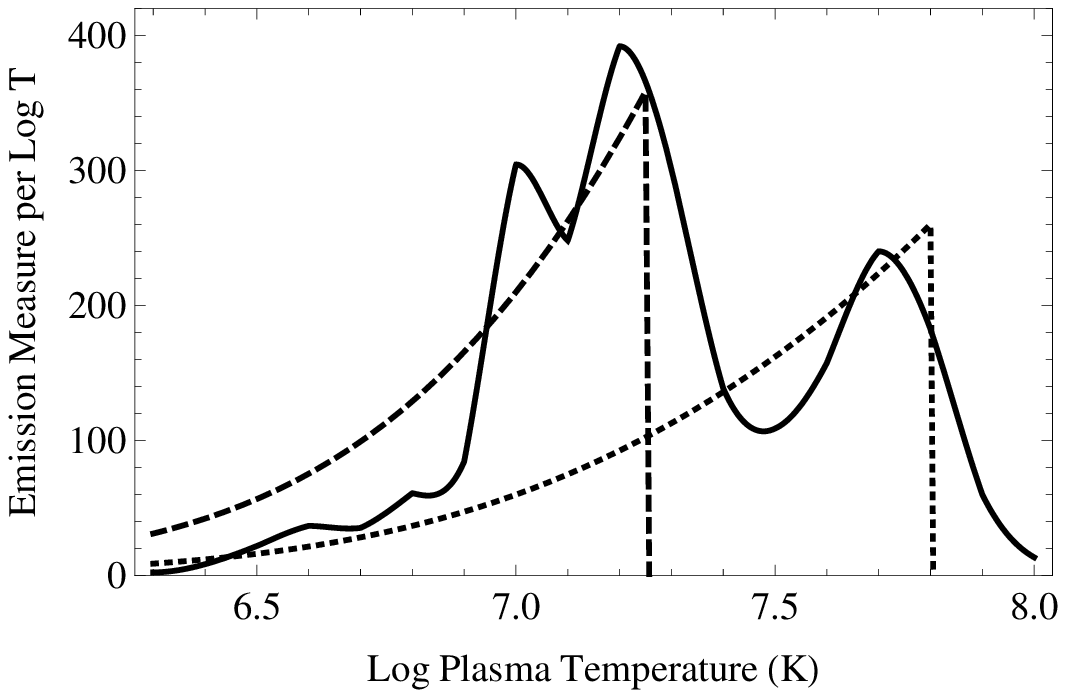}{2.6in}{-0.}{300.}{200.}{0}{0}
%\plotfiddle{PSFILE}{VSIZE}{ROTANG}{HSCALE}{VSCALE}{HTRANS}{VTRANS}
%\includegraphics[width=6.5in]{shockanglefig1.pdf}
\caption{
The solid curve is the DEM (per unit log $T$ rather than per unit $T$)
for $\sigma$ Gem inferred by Huenemoerder et al. (2013), and the dashed curves are the
DEM that are equivalent to two different single-initial-$T$ CIT models with initial $T = 18$ and 63 MK,
for pure radiative cooling using the heuristic approximations above.
}
\end{center}
\end{figure}
heated initial $T$ in the vicinity of 20 and 60 MK.
As with Figure 9, the bimodality can only be established if there are lines at the necessary $\ti$ points, and if
abundances do not create significant additional uncertainties about the DEM shape.
%%ref   Huenemoerder, D. P., Phillips, K. J. H., Sylwester, J., \& Sylwester, B. 2013, ApJ, 768, 135

\subsection{Hot-star winds}
%   --HSW (Zhekov)

Hot massive stars, including O and Wolf-Rayet stars, are often thermal X-ray sources (see Kuhn et al. 2013 for a recent update).
%%  Michael A. Kuhn, Konstantin V. Getman, Patrick S. Broos, Leisa K. Townsley, and Eric D. Feigelson  2013 ApJS 209 27
As these stars have dense high-velocity winds, it is believed the X-rays come from shocks, either within the unstable line-driven
winds (Owocki 1994; Lucy 2012; Sundqvist \& Owocki 2013), 
%%  Owocki, S. P. 1994, Ap\&SS, 221, 3
%%  Sundqvist, J. O. \& Owocki, S. P. 2013, MNRAS, 428, 1837
or in wind/wind collisions in hot-star binaries ().
In some cases, the thermal X-rays have been fit with a bimodal DEM, suggesting either the two wind shocks in the binary
interpretation, or a combination of a strong standing shock due to some kind of global wind interaction, and weaker stochastically
distributed internal shocks.
One such bimodal interpretation was given by Zhekov et al. (2011) 
%% ref Zhekov, S. A., Gagne, M., \& Skinner, S. L. 2011, ApJ, 727, 17
to the carbon-rich Wolf-Rayet star WR 48a, via their two-$T$ DEM fit.
They mention that the two $T$ might simply be proxies for the hotter and cooler end of a single process, such as a stronger shock near the axis
of a binary with weaker more oblique shocks away from that axis, or it might represent two spatially separate sources, such as
a triple system.
Still, the temperatures vary by only a factor of 2.7 for their non-ionization equilibrium (NEI) model, and 3.7 for their collisional
ionization equilibrium (CIE) model, and the above results suggest that such $T$ differences may be too narrow 
for bimodality to be unambiguously established.

For example, even taking the wider $T$ separations of the Zhekov et al. (2011) CIE model, 
using $\temp_1 = 0.87$ and $\temp_2 = 3.26$ with hotter/cooler emission
measure ratio of 1.2 as per that model, yields the heuristic comparison shown in Figure 13.
%% figure 13  zhekov 2T vs. 1S results or maybe power law
\begin{figure}[h]
\begin{center}
\plotfiddle{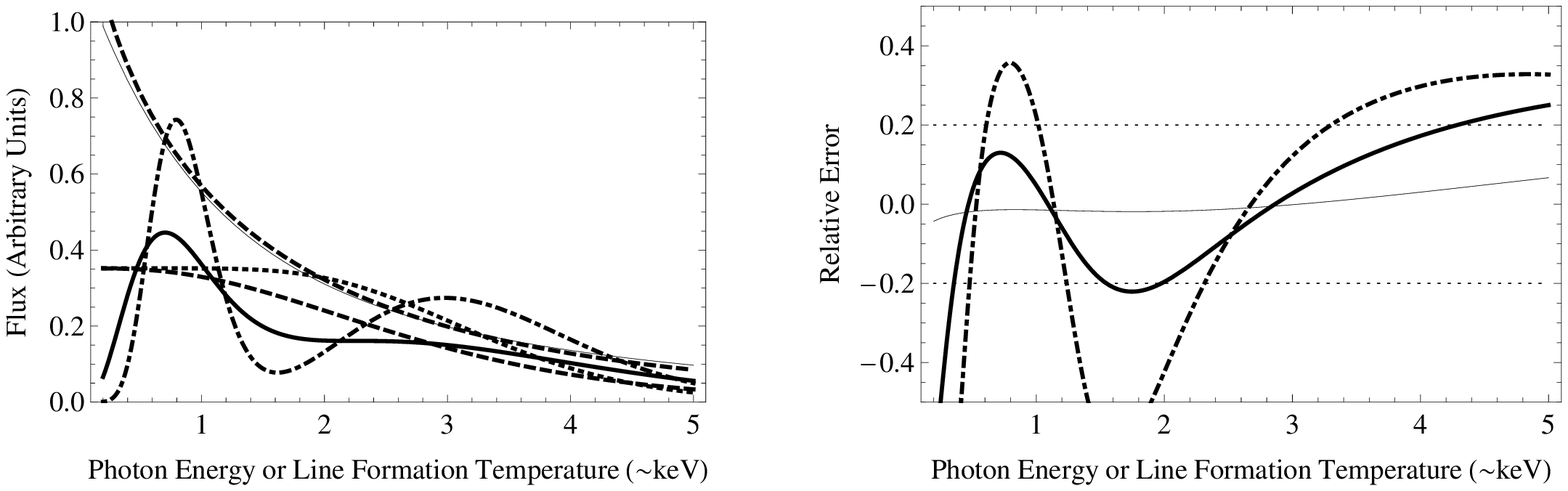}{2.6in}{-0.}{440.}{200.}{0}{0}
%\plotfiddle{PSFILE}{VSIZE}{ROTANG}{HSCALE}{VSCALE}{HTRANS}{VTRANS}
%\includegraphics[width=6.5in]{shockanglefig1.pdf}
\caption{
Left panel: The approximate bremsstrahlung continuum and heuristic
NLF distributions in arbitrary units, as a function of $\energy$ and $\ti$
respectively, comparing a two-$T$ DEM model with $\temp = 0.87$ and 3.26 and a hotter/cooler
emission-measure ratio of 1.2, with a single-initial-$T$ CIT model
with $\tinitial = 3.7$.
The curve conventions are as in Figs. 1--8, as is the right panel.
}
\end{center}
\end{figure}
The heuristic results for the bremsstrahlung continuum and the NLF distribution are compared to a single-initial-$T$ CIT model
with $\tinitial = 3.7$, corresponding to impulsive heating to about 3.7 keV and subsequent radiative cooling through all lower $T$.
Again we conclude that the bimodality of the
DEM approach is only evidenced in lines at the appropriate $\ti$, including $\ti < 0.5$, which is even less than
3/4 of the lower $\temp$ in the DEM model.
Given the potential for photoelectric absorption of lines with longer wavelengths, this opens the possibility that the inferred bimodality
is just a modeling choice and not an inherent attribute of the source, though this issue requires more careful investigation to resolve.
Investigations of the wind of Zeta Pup by Cohen et al. (2014) offer promise in distinguishing processes that maintain plasma continuously
at high $T$, such as might favor a DEM-type parametrization, from processes that impulsively heat gas which subsequently cools radiatively,
such as might favor a CIT-type parametrization.

\subsection{Cool-star coronae}
%   --coronae (Gudel)

For stars like the Sun, or younger versions of the Sun such as T Tauri stars, G{\"u}del, Guinan, \& Skinner (1997) looked at
many examples of such X-ray sources, and often found two-$T$ DEM fits.
For example, one of the more active T Tauri stars, EK Dra, was fit with the two-$T$ model that maps into the heuristic
analysis shown in Figure 14.
%% figure 14
\begin{figure}[h]
\begin{center}
\plotfiddle{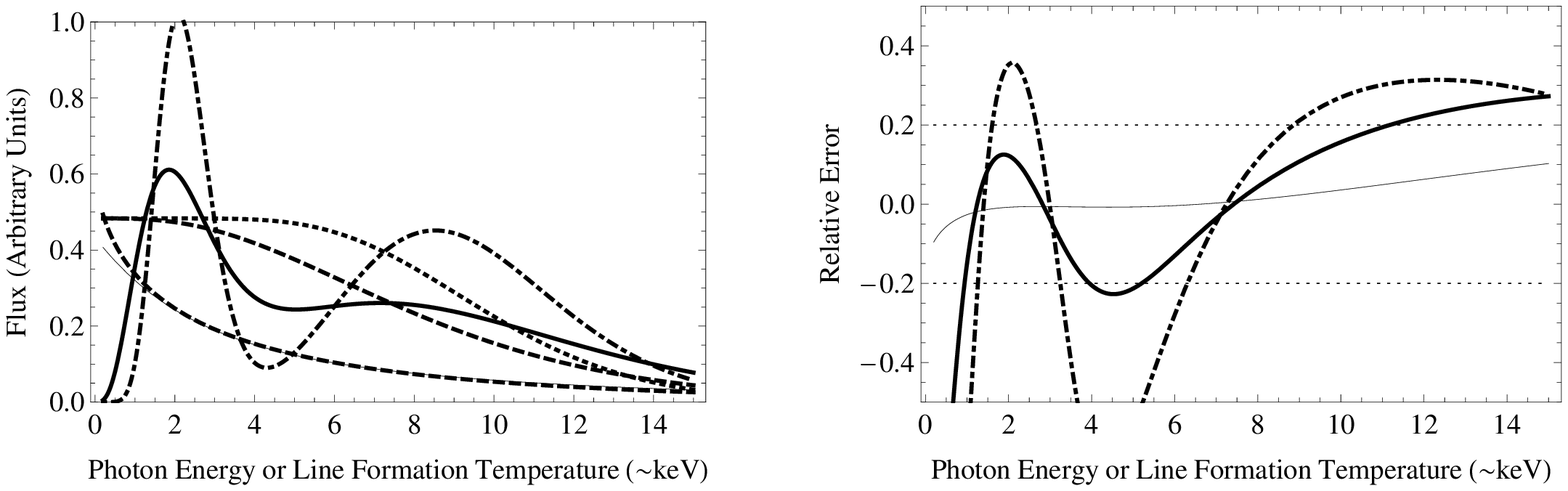}{2.6in}{-0.}{440.}{200.}{0}{0}
%\plotfiddle{PSFILE}{VSIZE}{ROTANG}{HSCALE}{VSCALE}{HTRANS}{VTRANS}
%\includegraphics[width=6.5in]{shockanglefig1.pdf}
\caption{
Left panel: The approximate bremsstrahlung continuum and heuristic
NLF distributions in arbitrary units, as a function of $\energy$ and $\ti$
respectively, comparing a two-$T$ DEM model with $\temp = 2.29$ and 9.93 and a hotter/cooler
emission-measure ratio of 1.16, with a single-initial-$T$ CIT model
with $\tinitial = 11$.
The curve conventions are as above, as is the right panel.
}
\end{center}
\end{figure}
The parameters are $\temp = 2.29$ and 9.93, again in units where $\To = 1$ keV, and the hot/cool emission measure ratio
is 1.16.
Figure 14 shows what has by now become a typical story, whereby a single-initial-$T$ CIT model can yield almost perfect
agreement to the bremsstrahlung continuum, here using $\tinitial = 11$, and then the NLF distribution also agrees at
many possible $\ti$ where lines might be located, though no good agreement is possible for $\ti < 1$, and lines with
narrow emissivities that happen to fall near $\ti = 2, 4$ or 9 could not fit both the two-$T$ DEM and a smooth CIT distribution.
However, specific lines that do not fit the NLF distribution could be brought into agreement if the abundances are regarded
as variable, so it is not clear without a more careful analysis if bimodality is a robust feature of the source model of EK Dra.
G{\"u}del et al. (1997) cite temporal variability as the primary way to distinguish the hotter and cooler components, as flaring
activity tends to create greater variability in the hotter component.
Note that temporal bimodality is a bit different from spatial bimodality, as even a CIT model with two
impulsive heating mechanisms could exhibit
temporal bimodality, yet still give a locally smooth DEM and a monotonic NLF distribution.
As usual, the main diagnostic that distinguishes continuously maintained DEM sources from impulsively heated and transiently
cooled CIT sources is the absence or presence of the cooling gas, so that is the issue that requires special attention in
terms of what low-$\ti$ lines are accessible, and what role is played by foreground absorption of long-wavelength lines.

It should be noted that heat conduction has not been expressly included in this simple picture, which is likely to be an
important omission given the success of conductive loop models in the solar corona.
Rather than focus on spatially and temporally stochastic structures without conduction, an alternative approach for obtaining
analytic results is pursued by Jordan, Ness, \& Sim (2012), who instead take the spatial and temporal structure as steady-state,
and concentrate on the DEM produced when radiative cooling induces a divergence in the conductive flux.
Most likely the true situation involves some kind of combination of these two idealizations, in relative measure that may
well differ from source to source of stellar X-rays (G{\"u}del \& Naze 2009).

\subsection{Supernova remnants}
%   --SNR (McEntaffer)

%%ref  McEntaffer, R. L. \& Brantseg, T. 2011, ApJ, 730, 99
%%ref  McEntaffer, R. L., Grieves, N., DeRoo, C., \& Brantseg, T. 2013, ApJ, 774, 120
%%  the second one brings in spatial resolution as an important supporting information to the spectral fits, indeed the fits
%%  often use two T that are too close to be considered bimodal
Another astrophysical application where two-$T$ DEM fits have proven successful is in supernova remnants.
For example, McEntaffer \& Brantseg (2011) find that X-ray emission from 
the Cygnus Loop can be spatially resolved into several different regions
with their own two-$T$ DEM fits.
The fit parameters vary over the sample, but one common attribute is that when there are two well-separated $T$ in their
fit, the hot gas has vastly less emission measure than the cool gas.
For example, their A1 region of of the Cygnus Loop 
is fit with $T = 0.086$ and 0.236 keV, with a hot/cool emission-measure ratio of 0.009,
so yields for $\To = 1$ keV the heuristic results shown in Figure 15.
%% figure 15
\begin{figure}[h]
\begin{center}
\plotfiddle{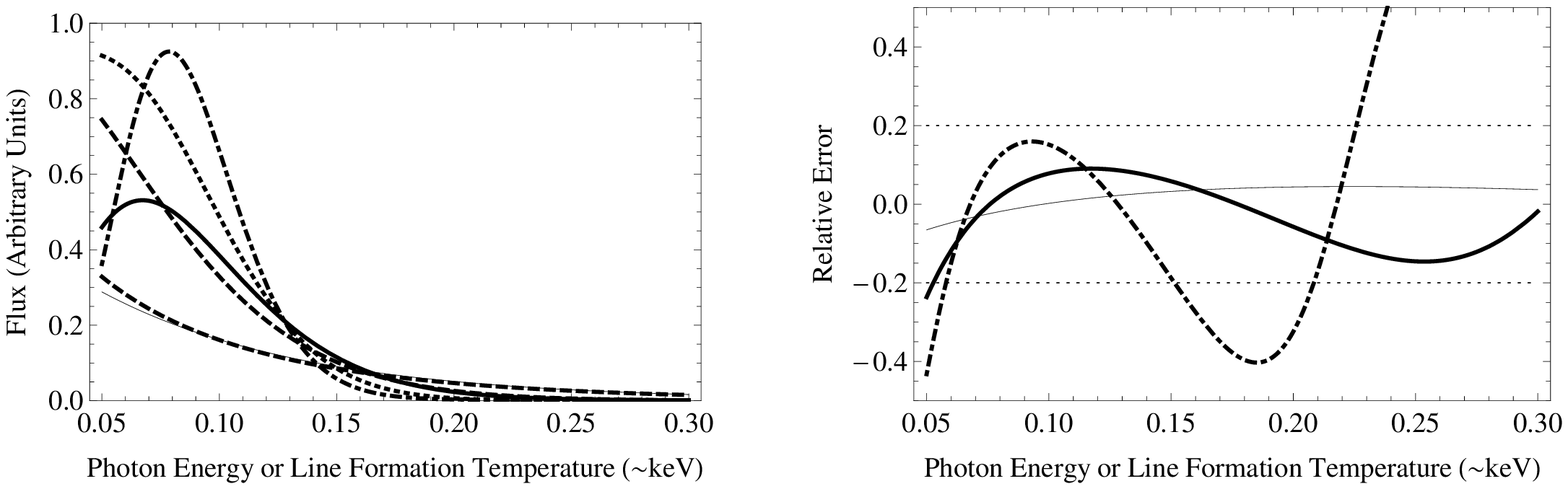}{2.6in}{-0.}{440.}{200.}{0}{0}
%\plotfiddle{PSFILE}{VSIZE}{ROTANG}{HSCALE}{VSCALE}{HTRANS}{VTRANS}
%\includegraphics[width=6.5in]{shockanglefig1.pdf}
\caption{
Left panel: The approximate bremsstrahlung continuum and heuristic
NLF distributions in arbitrary units, as a function of $\energy$ and $\ti$
respectively, comparing a two-$T$ DEM model with $\temp = 0.086$ and 0.236 and a hotter/cooler
emission-measure ratio of 0.009, with a single-initial-$T$ CIT model
with $\tinitial = 0.12$.
The curve conventions are as above, as is the right panel.
}
\end{center}
\end{figure}
The shown comparison is for a single-initial-$T$ model with $\tinitial = 0.12$, which fits an isothermal $\temp = 0.086$
NLF distribution except at low $\ti$.
The hard component in the two-$T$ DEM model does not make a significant impact on the NLF distribution because the
emission measure in that component is so low that it is difficult to see in the NLF, which suggests that it is
difficult from spectral analysis alone to isolate discrete bimodality.
However, spatially resolved spectra show $T$ differences that are much more unambiguously interpreted as discrete
components in a spatially inhomogeneous source distribution, as shown by McEntaffer et al. (2013) for the galactic
supernova remnant G272.2-3.2.
Hence, using complementary support from spatial resolution, the picture that emerges is more like multiple discrete
components that each have a single characteristic $T$, rather than components whose source distributions are
unambiguously locally bimodal.

\subsection{Discussion}

The above shows that a heuristic treatment of a new line-flux analysis tool, the NLF distribution, can lend insight
into more detailed efforts at spectral modeling.
Applying this to a wide array of phenomena that are sometimes fit with two-$T$ DEM models,
it is generally found that a simpler CIT model that involves impulsive heating to the
vicinity of a single initial $T$, followed by complete radiative cooling, can fit the entire bremsstrahlung continuum
essentially perfectly, and can also fit
lines at many different formation temperatures $\ti$, well enough to mimic a two-$T$ DEM fit under some circumstances.
The primary exceptions are at low $\ti$, where transiently cooling gas presents its signature, and at $\ti$ near 
any discrete features in the assumed DEM model.
Distinctions are more easily drawn using lines with a FWHM of about a factor of 2 in their emissivity functions,
whereas for the many lines where this factor is around 3, features in the DEM can only be discerned over coarser
$T$ scales.
As a general rule of thumb, lines can only distinguish structure in the DEM on $T$ scales coarser than the
emissivity function of the lines themselves, and whether or not gas is continuously maintained at some $T$ or 
allowed to cool is only distinguished by lines 
with peak formation temperatures 
less than about 1/2 or 3/4 of that $T$, for broader or narrower emissivity functions respectively.

Importantly, lines must have $\ti$ values at the appropriate points to be of value, as fitting lines with $\ti$
in regions of overlap between alternative models can yield undue confidence in the specifics of a particular model.
The above analysis can help ascertain what the important $\ti$ values are for any given DEM model, and whether or
not there are accessible lines with the appropriate $\ti$ that could build confidence in that model.
Since low-$\ti$ lines tend to appear at longer wavelengths where photoionization may truncate the line and continuum
fluxes, careful consideration must be given to the ambiguities that might introduce when the absorbing column
is not well constrained.
Finally, when abundances are also not well constrained, DEM features may become degenerate with changes in the abundances
of species with lines with $\ti$ at critical places in the DEM structure.
All of these issues suggest caution when inferring discrete DEM features such as bimodality, and a simplified analysis
such as the above can help navigate these uncertainties.
On the other hand, spatially resolved (such as in galactic nebulae) or temporally varying (such as flare stars)
structure in the DEM make it much easier to confidently establish separate or discrete source components,
without the need for this kind of careful spectral analysis.

\section{Photoelectric absorption corrections}
%% insert the concept of a column-mass-corrected NLF and bremsstrahlung continuum, based on the assumption that Ti depends on E
%%   and tau depends on E (see Cohen)
%%stopped here
%% MNRAS 375, 145 (2007)
%% Table 1 : cool model is 0.5e22 abs + 0.2 and 1 keV ; hot model is 0 abs + 0.6 and 2 keV

Many of the X-ray emissions outside our solar system must pass through significant photoelectric absorbing column prior
to observation, which can truncate the longer wavelength component because the cross sections typically rise at lower
energies closer to the ionization edges.
Given the importance of lines with low formation temperature $\ti$ in distinguishing a DEM with a minimum $T$ from a
cooling flow or other CIT-type source, and the fact that lower-$\ti$ lines often appear at longer wavelengths, the 
possibility for intervening absorption to squelch the emissions from potentially cooling gas is especially problematic.
Let us extend the above heuristic analysis to explore the possible impact of foreground absorption by generating
absorption-corrected NLF distributions, and absorbed bremsstrahlung continua.

A rough approximation for schematic results is that the optical depth decreases with photon energy
according to the canonical $\energy^{-2}$, although 
in practice the accumulation of photoionization edges can weaken the $\energy$ dependence, so the details vary
with the conditions in each astrophysical environment (e.g., for general effects, e.g. Starace 1982; for
%% Starace, A. F. 1982, HDP, 31, 1
hot-star winds and the ISM, e.g. Oskinova et al. 2003 and 
Leutenegger et al. 2010; for important oxygen photoabsorption, e.g. Gorczycka et al. 2013).
%%  Gorczycka, T. W., Bautista, M. A., Hasoglu, M. F., Garcia, J., Gatuzz, E., Kaastra, J. S., Kallman, T. R., 
%%  Manson, S. T., Mendoza, C., Raassen, A. J. J., De Vries, C. P., \& Zatsarinny, O. 2013, ApJ, 779, 78
Also, fitting to Huenemorder et al. (2013), a rough connection between the $\ti$ of the lines and their $\energy$,
for lines other than the L-shell Fe lines that typically have low $\energy$ for their $\ti$,
is given by
\beq
\label{trend}
\energy \ \cong \ \sqrt{3} \ti^{2/3} \ .
\eeq
Combining this with the $\energy$ dependence of the absorbing optical depth $\tau \propto \energy^{-2}$ then gives
\beq
\tau \ \cong \ \left ( \frac{\energy_1}{\energy} \right )^2 \ \cong \ \frac{1}{3} \energy_1^2 \ti^{-4/3} \ ,
\eeq
where $\energy_1$ is the photon energy (here in $k\To$ units, but roughy in keV) where the absorbing screen
reaches optical depth unity.
Then $e^{-\tau}$ multiplies the NLF distribution and the bremsstrahlung continuum, where $\tau$ is taken as a function
of $\ti$ for the former and of $\energy$ for the latter, to account schematically for absorption effects.

Applying this approach to the isothermal versus single-initial-$T$ comparison from Figure 1 yields the results
in Figure 16, with $\temp = 1.7$ and $\tinitial = 2.6$, where the optical depth is
characterized by $\energy_1 = 0.2$ as a generic value of interest, so the absorbing column has optical depth
unity at photon energies of about 0.2 keV in this example.
%% figure 16
\begin{figure}[h]
\begin{center}
\plotfiddle{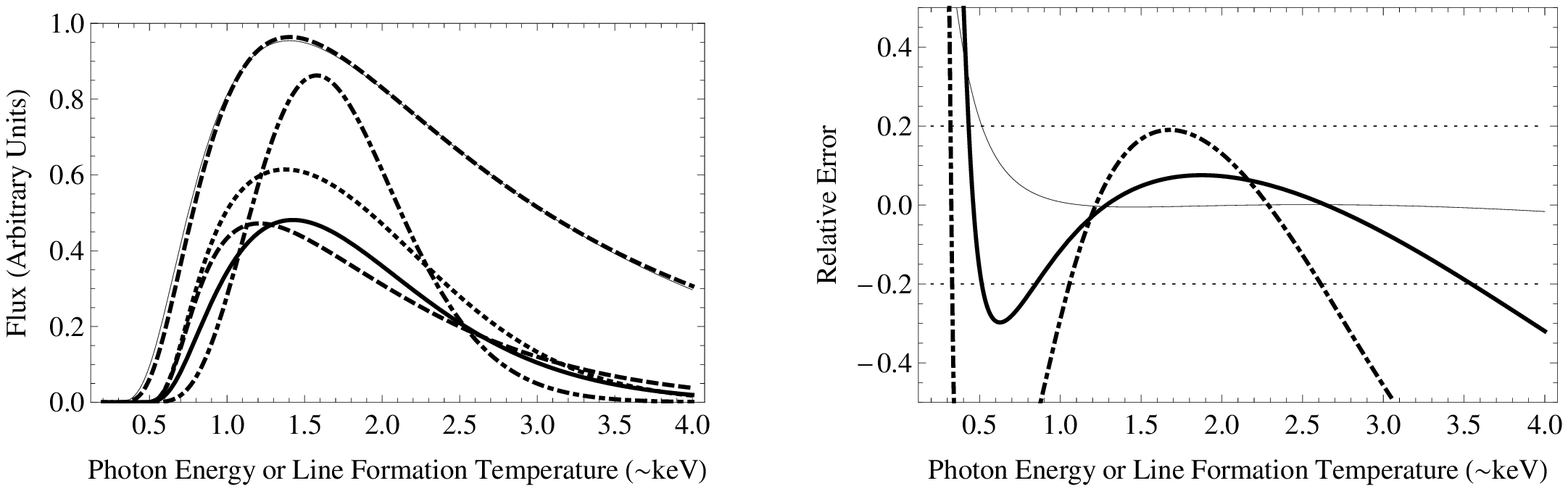}{2.6in}{-0.}{440.}{200.}{0}{0}
%\plotfiddle{PSFILE}{VSIZE}{ROTANG}{HSCALE}{VSCALE}{HTRANS}{VTRANS}
%\includegraphics[width=6.5in]{shockanglefig1.pdf}
\caption{
Left panel: The approximate bremsstrahlung continuum and heuristic
NLF distributions in arbitrary units, as a function of $\energy$ and $\ti$
respectively, comparing an isothermal DEM model with $\temp = 1.7$ 
with a single-initial-$T$ CIT model
with $\tinitial = 2.6$, where all curves are corrected for the simplified photoelectric absorption
described in the text, and the curve conventions are as above.
In effect, the theoretical curves are being corrected for absorption, rather than removing
the absorption from the observed fluxes.
The right panel shows the difference divided by the average for each comparison, as above.
}
\end{center}
\end{figure}
The figure shows that indeed the potential for identifying the presence of cooling gas 
is significantly obfuscated by the absorption, unless lines can be found with low $\ti$ but unusually high $\energy$
so as not to fit the general trend in eq. (\ref{trend}).
It also shows that the poor fit between single-$T$ DEM and single-initial-$T$ CIT models can be significantly
improved if absorption takes away much of the distinguishing $\ti$ regime (for the NLF distribution)
and $\energy$ regime (for the continuum).

Similar ambiguities owing to foreground absorption are noted by Naze et al. (2007),
%% ref Naze 
who find that uncertainties in the foreground absorption can create confusion between harder sources, 
and softer sources with significant absorption of the cooler component.
Figure 17
%% figure 17
\begin{figure}[h]
\begin{center}
\plotfiddle{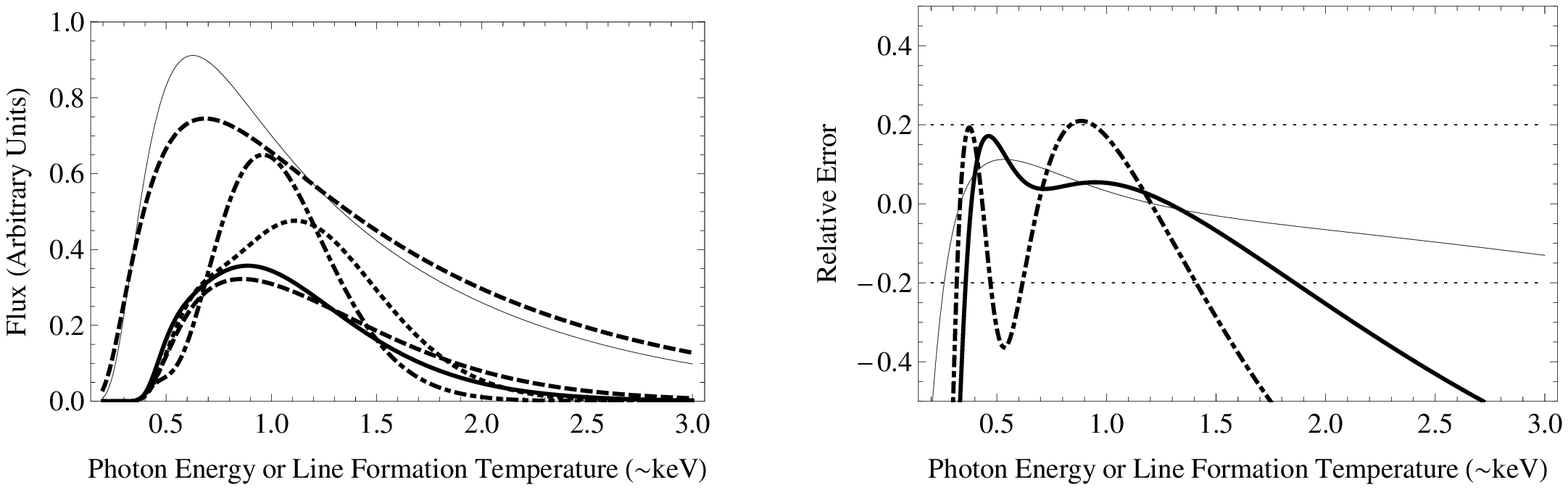}{2.6in}{-0.}{440.}{200.}{0}{0}
%\plotfiddle{PSFILE}{VSIZE}{ROTANG}{HSCALE}{VSCALE}{HTRANS}{VTRANS}
%\includegraphics[width=6.5in]{shockanglefig1.pdf}
\caption{
Left panel: The approximate bremsstrahlung continuum and heuristic
NLF distributions in arbitrary units, as a function of $\energy$ and $\ti$
respectively, comparing two different two-$T$ DEM models with $\temp = 0.27$ and 1.01, and
0.60 and 1.23 for the other model,
where all curves are corrected for the necessary amount of photoelectric absorption
needed to make them roughly agree.
The hotter/cooler emission-measure ratios for the two models are 0.567 and 2.97, and the values of
$\energy_1$ in the two models are 0.5 and 0.4, all to produce a schematic rendition of the
mutual fits presented in Naze et al. (2007).
The curve conventions are as above.
Right panel: The same curves, except showing the comparison of the difference divided by
the average for both two-$T$ DEM models, with the usual curve conventions.
}
\end{center}
\end{figure}
shows how the heuristic analysis from this paper could be used to make a similar point, by taking the
two-$T$ models from Naze et al. (2007) with $\temp = 0.27$ and 1.01 in one model, and 0.60 and 1.23 in the other,
all assuming $\To = 1$ keV, and hotter/cooler emission-measure ratios of 0.567 and 2.97 in the two models respectively.
The figure treats the total emission measure, and the degree of foreground absorption, as adjustable variables,
and corroborates Naze et al. (2007) that the two models can be difficult to distinguish.
This suggests the presence of degeneracies between  quite different
two-$T$ DEM models, let alone other types of source models like the CIT approaches focused on here.

%%  first give my own fits 2t and 1s with absorption, showing the cooling gas is left ambiguous, then 
%%  cite how Naze found even a 2t fit can be ambiguous with a softer 2t fit

\section{Conclusions}

This paper defines a new line-flux diagnostic, the normalized line-flux (NLF) distribution, as a function of formation temperature $\ti$,
and applies a heuristic analytic treatment of the flux kernels to explore how different source parametrizations yield different observable
line and continuum diagnostics.
The primary goal was to assess the degree to which thermal X-ray spectra can be said to be fitted in a model-independent way, rather than
leaving the signature of the choices of the modeler in ways that could unduly skew the interpretation of the results.
It was found that significant ambiguity does exist, such that true model independence is difficult to establish.
This further suggests that the modeler may be better served by embracing a parametrization that connects directly with an interesting
type of heating physics, such as impulsive heating followed by transient cooling as in the cumulative initial temperature (CIT) approach.
When the transient cooling is predominantly radiative, parametrizing a CIT fit then gives direct constraints on the rate that gas is
being raised to various different initial temperatures, whether by shocks or flaring of some kind.

Even though the CIT approach and the DEM approach are formally equivalent in the sense that one can be translated into the other
via eq. (\ref{conversion}), the value in the CIT approach is that it is constrained to follow a monotonically decreasing source distribution, and
requires an NLF distribution and a bremsstrahlung continuum that
are also monotonically decreasing, after photoelectric absorption corrections.
This makes it relatively easy, at least in principle, to tell if a CIT-type model will be appropriate, though abundance uncertainties
can make it difficult in practice to establish unambiguously if the NLF distribution is truly monotonic when using different species
in different $\temp$ regimes.
As such, it may be better to embrace a particular heating hypothesis,
such as a CIT approach, and explore the abundance and absorption requirements necessary to justify that
interpretation, thereby capitalizing on the organizing features of model {\it dependence.}
This may allow more useful constraints to be extracted from the observations, than pursuing 
the opposite approach of attempting complete model independence, which is likely impossible due to the ambiguities introduced
when trying to use the data to answer too many questions at once.

Throughout the process of arriving at these conclusions, a number of more specific conclusions were also arrived at.
These include:
\begin{enumerate}
\item The bremsstrahlung continuum can be fit remarkably precisely by completely different source models, despite the
formal invertibility of the Laplace transform, owing to its ill conditioning.
\item Single-initial-$T$ CIT approaches can yield an NLF distribution that
fits a two-$T$ DEM if sufficiently low-$\ti$ lines and other well-placed lines are not accessible, perhaps because of foreground absorption,
uncertain abundances in key species, or a lack of lines with narrow emissivities at the necessary formation temperatures.
\item A single-initial-$T$ CIT involves only 1 shape parameter, whereas a two-$T$ DEM invokes 3, so the former is always a simpler
fit whenever it is successful, and may provide a better starting point for additional fitting adjustments any time impulsive heating
is inferred or advocated.
\item Since line emissivity functions typically exhibit a factor 2--3 FWHM in $\ti$, features in the DEM on similar
$T$ scales cannot be unambiguously inferred, especially if abundances are uncertain.
\item The key discriminant of a radiatively-cooled CIT model is the signature of cool gas, whereas a discrete DEM has no gas cooler than some $T$,
so abundance and absorption uncertainties must be navigated to assess whether or not cool gas is actually present.
\item Non-radiative cooling processes that contribute to the $f_o(\temp)$ term in eq. (\ref{scpt}) can also mimic the absence of cool gas,
but analysis of the impact of $f_o$ is beyond the current scope of this analysis.
\item A power-law CIT parametrization can improve the fit to a two-$T$ DEM using 2 shape parameters instead of 3, and can be useful when
line fluxes tail off gradually at short wavelengths.
\item Many astrophysical applications that choose two-$T$ DEM fits could possibly also be fit with a single-initial-$T$ CIT whenever there exists
sufficient motivation for doing so, if observational sensitivity
or uncertainties in abundances or absorption make the absence of low-$\ti$ lines difficult to establish unambiguously.
\item Discrete DEM fits cannot be considered model independent 
without careful consideration of whether or
not the discreteness is a justifiably robust feature of any successful fit.
\item Discrete emission components can be established spatially or temporally more easily than with spectral analysis.
\end{enumerate}

These conclusions were reached by considering a wide variety of two-$T$ DEM fits that span a host of astrophysical contexts, from
the scale of stars to galaxy clusters.
Many authors that use two-$T$ fits do not interpret them as literally discrete source components, while others sometimes do.
Yet it is striking that so many different observations were successfully fit with two-$T$ fits of a rather similar character,
and it seems unlikely that so many different X-ray generating mechanisms would be fundamentally discrete or bimodal.
Although exceptions exist, many of the fits involved a higher $T$ that was roughly a factor of 3 above the lower $T$, which
is found by the above analysis to be essentially the highest ratio whereby typical line emissivity functions can satisfy a
bimodal interpretation even when the plasma itself is not bimodal.
True bimodality in the DEM at the level of a factor of 3 in $T$ can 
only be inferred if lines exist at the necessary $\ti$, if their abundances are well
constrained, and if the lines with narrower $T$ precision are stressed, which
are not necessarily the strongest lines or the lines with highest signal-to-noise.  

Note that a factor of 3 in $T$ allows emission from lines over an order of magnitude of different 
formation $T$, given the breadth of the line
emissivity functions, which means it is capable of
mimicking a wide actual source distribution.
Smaller ratios would not span as wide a $T$ range, so differ little from isothermal fits,
whereas larger ratios would require that the observations show a more obvious and 
unambiguous bimodality in their NLF distribution, so are more rarely encountered in the context of modern high-resolution data.
It is suggested that this state of affairs is not a coincidence, but rather, it may just be the type of two-$T$ fit one obtains if one
attempts to match the spectrum from a smooth DEM with a two-$T$ DEM.
Hence, although a two-$T$ DEM fit does successfully characterize general attributes of the DEM such as average $T$ and
range of different $T$, it cannot be taken as evidence of discrete source components, without further supporting evidence
from a more careful investigation.
That more careful investigation can include spatial and temporal decomposition, and/or reliable diagnostics of the presence or
absence of a continuous chain of cooler $T$ than the minimum $T$ required for a successful discrete DEM fit.
The significance of this difference is that it implies a totally different heating mechanism, one that heats the gas
impulsively rather than maintaining it locally in a steady state.

The author would like to acknowledge helpful discussions with Maurice Leutenegger, Randall McEntaffer, Philip Judge,
David Huenemoerder, Yael Naze, and David Cohen.
This work was supported by NASA grant NNX11AF83G to the University of Iowa and NSF grant GO3-13003C to the 
Smithsonian Institution.

\end{document}